\documentstyle[epsfig,mathptm]{mn}
\newif\ifAMStwofonts

\def\nd#1#2{{d #1 \over d #2}}

\def\tfrac#1#2{{\textstyle\frac{#1}{#2}}}
\def\vect#1{{\mathbfit{#1}}}

\ifoldfss
  \ifCUPmtlplainloaded \else
    \NewTextAlphabet{textbfit} {cmbxti10} {}
    \NewTextAlphabet{textbfss} {cmssbx10} {}
    \NewMathAlphabet{mathbfit} {cmbxti10} {} 
    \NewMathAlphabet{mathbfss} {cmssbx10} {} 
  \fi
  \ifAMStwofonts
    \ifCUPmtlplainloaded \else
      \NewSymbolFont{upmath} {eurm10}
      \NewSymbolFont{AMSa} {msam10}
      \NewMathSymbol{\upi}     {0}{upmath}{19}
      \NewMathSymbol{\umu}     {0}{upmath}{16}
      \NewMathSymbol{\upartial}{0}{upmath}{40}
      \NewMathSymbol{\leqslant}{3}{AMSa}{36}
      \NewMathSymbol{\geqslant}{3}{AMSa}{3E}

    \fi
  \fi
\fi 

\ifnfssone
  \newmathalphabet{\mathit}
  \addtoversion{normal}{\mathit}{cmr}{m}{it}
  \addtoversion{bold}{\mathit}{cmr}{bx}{it}
  \newmathalphabet{\mathbfit} 
  \addtoversion{normal}{\mathbfit}{cmr}{bx}{it}
  \addtoversion{bold}{\mathbfit}{cmr}{bx}{it}
  \newmathalphabet{\mathbfss} 
  \addtoversion{normal}{\mathbfss}{cmss}{bx}{n}
  \addtoversion{bold}{\mathbfss}{cmss}{bx}{n}
  \ifAMStwofonts
    \ifCUPmtlplainloaded \else
      \UseAMStwoboldmath
      \makeatletter
      \new@mathgroup\upmath@group
      \define@mathgroup\mv@normal\upmath@group{eur}{m}{n}
      \define@mathgroup\mv@bold\upmath@group{eur}{b}{n}
      \edef\UPM{\hexnumber\upmath@group}
      \new@mathgroup\amsa@group
      \define@mathgroup\mv@normal\amsa@group{msa}{m}{n}
      \define@mathgroup\mv@bold\amsa@group{msa}{m}{n}
      \edef\AMSa{\hexnumber\amsa@group}
      \makeatother
      \mathchardef\upi="0\UPM19
      \mathchardef\umu="0\UPM16
      \mathchardef\upartial="0\UPM40
      \mathchardef\leqslant="3\AMSa36
      \mathchardef\geqslant="3\AMSa3E
    \fi
  \fi
\fi 

\ifnfsstwo
  \DeclareMathAlphabet{\mathbfit}{OT1}{cmr}{bx}{it}
  \DeclareMathAlphabet{\mathbfss}{OT1}{cmss}{bx}{n}
  \ifAMStwofonts
    \ifCUPmtlplainloaded \else
      \DeclareSymbolFont{UPM}{U}{eur}{m}{n}
      \SetSymbolFont{UPM}{bold}{U}{eur}{b}{n}
      \DeclareSymbolFont{AMSa}{U}{msa}{m}{n}
      \DeclareMathSymbol{\upi}{0}{UPM}{"19}
      \DeclareMathSymbol{\umu}{0}{UPM}{"16}
      \DeclareMathSymbol{\upartial}{0}{UPM}{"40}
      \DeclareMathSymbol{\leqslant}{3}{AMSa}{"36}
      \DeclareMathSymbol{\geqslant}{3}{AMSa}{"3E}
    \fi
  \fi
\fi 

\ifCUPmtlplainloaded \else
  \ifAMStwofonts \else 
    \def\upi{\pi}
    \def\umu{\mu}
    \def\upartial{\partial}
  \fi
\fi

\title
[Bayesian detection of discrete objects]
{A Bayesian approach to discrete object detection in astronomical datasets}
\author[M.P.~Hobson and C.~McLachlan]
{M.P.~Hobson and C.~McLachlan\\
Astrophysics Group, Cavendish Laboratory, Madingley Road, 
Cambridge, CB3 0HE, UK}
\date{Accepted ---. Received ---; in original form 26 April 2002}
\pagerange{\pageref{firstpage}--\pageref{lastpage}}
\pubyear{2002}

\begin{document}
\maketitle
\label{firstpage}

\begin{abstract}
A Bayesian approach is presented for detecting and characterising the
signal from discrete objects embedded in a diffuse background.  The
approach centres around the evaluation of the posterior distribution
for the parameters of the discrete objects, given the observed data,
and defines the theoretically-optimal procedure for parametrised object
detection. Two alternative strategies are investigated: the
simultaneous detection of all the discrete objects in the dataset, and
the iterative detection of objects. In both cases, the parameter space
characterising the object(s) is explored using Markov-Chain
Monte-Carlo sampling. For the iterative detection of objects, another
approach is to locate the global maximum of the posterior at each
iteration using a simulated annealing downhill simplex algorithm. The
techniques are applied to a two-dimensional toy problem consisting of
Gaussian objects embedded in uncorrelated pixel noise.  A cosmological
illustration of the iterative approach is also presented, in which the
thermal and kinetic Sunyaev-Zel'dovich effects from clusters of
galaxies are detected in microwave maps dominated by emission from
primordial cosmic microwave background anisotropies.
\end{abstract}

\begin{keywords}
cosmic microwave background -- methods: data analysis -- methods:
statistical.  
\end{keywords}

\section{Introduction}
\label{intro}

The detection and characterisation of discrete objects is a generic
problem in many areas of astrophysics and cosmology. Indeed, one of
the major challenges in the analysis of one-dimensional spectra,
two-dimensional images or higher-dimensional datasets is to separate a
localised signal from a diffuse
background. Typical one-dimensional examples include the extraction of
point or extended sources from time-ordered scan data or the detection
of absorption or emission lines in quasar spectra.  In two dimensions,
one often wishes to detect point or extended sources in astrophysical
images that are dominated either by instrumental noise or
contaminating diffuse emission. Similarly, in three dimensions, one
might wish to detect galaxy clusters in large-scale structure surveys.

To illustrate our discussion, we will focus on the important specific example
of detecting discrete objects in a diffuse background in
a two-dimensional astronomical image, although our general approach
will be applicable to datasets of arbitrary dimensionality.
Several packages exist
for performing this task, such as DAOfind (Stetson 1992), for
identifying stellar objects, and SExtractor (Bertin \& Arnouts 1996).
As pointed out by Sanz et al. (2001), when trying to detect discrete
sources, it is necessary to take proper account of the behaviour of 
the background emission, which typically contains contributions 
from astrophysical `contaminants' and instrumental noise.
It is often assumed (in both DAOfind and SExtractor) that 
the background is smoothly varying and has a characteristic scale 
length much larger than the scale of the discrete objects being
sought. For example, SExtractor approximates the background
emission by a low-order polynomial, which is subtracted from the
image. Object detection is then performed
by finding sets of connected pixels above some given threshold.

It is not uncommon, however, for astrophysical images to contain 
contaminating background emission that varies on length scales 
and with amplitudes close to those of the discrete objects of interest.
Moreover, the rms level of instrumental noise on the image may be
comparable to, or somewhat larger than, 
the amplitude of the localised signal one is seeking.
A specific example is provided by high-resolution observations
of the cosmic microwave background (CMB). In addition to the CMB
emission, which varies on a characteristic scale of order $\sim 10$
arcmin, one is often interested in detecting emission from
discrete objects such as extragalactic `point' (i.e. beam-shaped)
sources or the Sunyaev-Zel'dovich (SZ) effect in galaxy clusters, 
which have characteristic scales similar to that of the primordial
CMB emission. Moreover, the rms of the instrumental noise in 
CMB observations is often greater than the amplitude of the discrete
sources. In such cases, it is not surprising that straightforward 
methods, such as those outlined above, often fail to
detect the localised objects.

The traditional approach for dealing with such difficulties is 
to apply a linear filter to the original 
image $d(\vect{x})$ and instead analyse the resulting
filtered field $d_f(\vect{x})$. This process is usually performed
by Fourier transforming the image $d(\vect{x})$ to obtain
$\tilde{d}(\vect{k})$, multiplying by some filter function 
$\tilde\psi(\vect{k})$ 
and inverse Fourier
transforming. The form of the filter function clearly determines which 
Fourier modes are
suppressed and 
by what factor. In the simplest cases 
$\tilde\psi(\vect{k})$ may set to zero the amplitudes of
all $\vect{k}$-modes with $|\vect{k}|$ above (or below) some critical value 
$k_{\rm c}$, and one obtains a simple low-pass (or high-pass) Fourier
filter. Alternatively, one may retain only some specific set of Fourier modes
by setting to zero all modes with $|\vect{k}|$ 
lying outside some range $k_{\rm min}$ to $k_{\rm max}$ (see, for
example, Chiang et al. 2002). 
Clearly, the filtering procedure may also be considered 
as convolving the original image $d(\vect{x})$ with the function
$\psi(\vect{x})$ to obtain the filtered image
\begin{equation}
d_f(\vect{x}) = \int \psi(\vect{x}-\vect{y}) d(\vect{y})\,d^2\vect{y}.
\label{dfilt}
\end{equation}

Suppose one is interested in detecting objects
with some given spatial template $t(\vect{x})$ (normalised for
convenience to unit peak amplitude). If the original image contains
$N_{\rm obj}$ objects at positions $\vect{X}_i$ with amplitudes $A_i$, together with
contributions from  other astrophysical components and instrumental
noise, we may write
\[
d(\vect{x}) \equiv s(\vect{x})+n(\vect{x}) 
= \sum_{i=1}^{N_{\rm obj}} A_i t(\vect{x}-\vect{X}_i) + n(\vect{x}),
\]
where $s(\vect{x})$ is the signal of interest and $n(\vect{x})$ is
the generalised background `noise', defined as all contributions to
the image aside from the discrete objects. As shown by Sanz et al. (2001),
it is straightforward to design an optimal filter function
$\psi(\vect{x})$ such that the filtered field 
(\ref{dfilt}) has the following properties: (i) $d_f(\vect{X}_i)$ is an
unbiassed estimator of $A_i$; (ii) the variance of the filtered noise field  
$n_f(\vect{x})$ is minimised; (iii) $d_f(\vect{x})$ has local maxima
at the positions $\vect{X}_i$ of the objects. Sanz et al. 
call the corresponding function $\psi(\vect{x})$ the optimal pseudofilter.
In fact, it is closely related to the standard matched filter (see,
for example, Haehnelt \& Tegmark 1996) which is defined simply by
removing the condition (iii) above. In either case, one may consider
the filtering process as `optimally boosting' (in a linear sense) 
the signal from discrete 
objects, with a given spatial template, and simultaneously suppressing
emission from the background.

An additional subtletly in most practical applications is that the set
of objects are not all identical. Nevertheless, one can repeat the
filtering process with filter functions optimised for different
spatial templates to obtain several filtered fields, each of which will
optimally boost objects with that template. In the case of the SZ
effect, for example, one might assume that the functional form of the
template is the same for all clusters, but that the `core radius'
differs from one cluster to another. The functional form of the filter
function is then the same in each case, and one can repeat the
filtering for a number of different scales (Herranz et al. 2002a).

It is, of course, unnecessary to restrict our attention to a single
astronomical image, such as a two-dimensional map at a particular observing
frequency. In many cases, several different images may be available,
each of which contain information regarding the discrete objects of
interest. Once again, the SZ effect provides a good
example. Forthcoming CMB satellite missions, such as the Planck
experiment, should provide high-sensitivity, high-resolution
observations of the whole sky at a number of different observing
frequencies. Owing to the distinctive frequency dependence of the
thermal SZ effect, it is better to use the maps at all the observed
frequencies simultaneously when attempting to detect and characterise
thermal SZ clusters hidden in the emission from other astrophysical
components. The generalisation of the above filter
techniques to multifrequency data is straightforward, and leads to
the concept of multifilters (Herranz et al. 2002b). We also note that
an alternative approach to this problem, which relies only on the 
well-known frequency dependencies
of the thermal SZ and CMB emission, has been proposed by Diego et
al. (2001).

Although the approaches outlined above have been shown to produce
good results, the rationale for performing object detection in this
way is far from clear. Firstly, the distinction drawn between the 
filtering and object-detection steps is somewhat arbitrary. Also, the
filtering process itself is only optimal among the rather limited class
of linear filters. Therefore, in this paper, we present a general, 
non-linear, Bayesian approach to
the detection and characterisation of discrete objects in a diffuse
background. As in the filtering techniques, the method
assumes a parameterised form for the objects of interest, but the
optimal values of these parameters, and their associated errors, are
obtained in a single step by evaluating their
full posterior distribution. If available, one may also place physical
priors on the parameters defining an object and on the number of
objects present. The approach represents the theoretically-optimal
method for performing parametrised object detection. Moreover,
owing to its general nature, the
method may be applied to a wide range astronomical datasets.

The outline of the paper is as follows. In section~\ref{bayesinf}, 
we review the basic aspects of the Bayesian approach to parameter
estimation and the r\^ole of the Bayesian evidence in model selection.
In section~\ref{MCMCsampling} we discuss the
use of Markov-Chain Monte-Carlo (MCMC) methods in the implementation
of the Bayesian approach to inference.
In section~\ref{bayesobj}, we formulate the mathematical problem of 
detecting discrete objects in a background using Bayes theorem, 
and introduce a toy problem to illustrate the process. In
Section~\ref{simultaneous}, 
we present a technique for the simultaneous detection
of all the discrete objects in an image, whereas in Section~\ref{iterative} we
discuss two alternative iterative approaches in which objects are
identified one-by-one. We illustrate the iterative approach
to Bayesian object detection
in section~\ref{szdetect}, by applying it to the
interesting problem of identifying the SZ effect from galaxy clusters 
in microwave maps dominated by primordial CMB emission.
Finally, our conclusions are presented in section~\ref{conc}.

\section{Bayesian inference}
\label{bayesinf}

We begin by reviewing briefly the basic principles of Bayesian
inference. For some dataset $\vect{D}$, suppose we are
interested in estimating the values of a set of parameters
$\btheta$ in some underlying model of the data.
For any given model, one may write down an expression for the 
{\em likelihood} $\Pr(\vect{D}|\btheta)$ of obtaining the data 
vector $\vect{D}$ given a particular
set of values for the parameters $\btheta$. In addition to the
likelihood function, one may impose a {\em prior} $\Pr(\btheta)$ on the
parameters, which represents our state of knowledge (or prejudices)
regarding the values of the parameters {\em before} analysing the data
$\vect{D}$. Bayes' theorem then reads,
\begin{equation}
\Pr(\btheta|\vect{D})=\frac{\Pr(\vect{D}|\btheta)\Pr(\btheta)}
{\Pr(\vect{D})},
\label{bayes}
\end{equation}
which gives the {\em posterior} distribution $\Pr(\btheta|\vect{D})$ 
in terms of the likelihood, the prior and the {\em evidence} 
$\Pr(\vect{D})$ (which is also often called the marginalised likelihood). 

\subsection{Parameter estimation}

For the purpose of estimating parameters, one usually
ignores the normalisation factor $\Pr(\vect{D})$ in Bayes' theorem,
since it does not depend on the values of the parameters $\btheta$.
Thus, one normally works instead with the `unnormalised posterior'
\begin{equation}
\overline{\Pr}(\btheta|\vect{D}) \equiv
\Pr(\vect{D}|\btheta)\Pr(\btheta),
\label{pbardef}
\end{equation}
where we have written $\overline{\Pr}$ to denote the fact that 
the `probability distribution' on the left-hand side is
not normalised to unit volume. In fact, it is also common to omit
normalising factors, that do not depend on the parameters $\btheta$,
from the likelihood and the prior. As we shall see below, however, 
if one wishes to calculate the Bayesian evidence for a particular
model of the data, the likelihood and the prior must be properly normalised
such that $\int\Pr(\btheta)\,d\btheta=1$ and
$\int\Pr(\vect{D}|\btheta)\,d\vect{D}=1$. We will therefore 
assume here that the necessary normalising factors have been retained.

Strictly speaking, the {\em entire} (unnormalised) 
posterior is the Bayesian inference
of the parameters values. Ideally, one would therefore wish
to calculate $\overline{\Pr}(\btheta|\vect{D})$ throughout some 
(large) hypercube in parameter space.
Unfortunately, if the dimension of the parameter space is large, this
is often numerically unfeasible. Thus, particularly in large problems,  
it has been common practice 
to  present one's results in terms of the `best' estimates
$\hat{\btheta}$, which maximise the (unnormalised) posterior, 
together with some associated errors, usually
quoted in terms of the estimated covariance matrix
\begin{equation}
\vect{C} = \left[-\left.\nabla\nabla\ln\overline{\Pr}(\btheta|\vect{D})
\right|_{\btheta=\hat{\btheta}}\right]^{-1}.
\label{covmatdef}
\end{equation}

The estimates $\hat{\btheta}$ are usually obtained by 
an iterative numerical minimisation algorithm. 
Indeed, standard numerical algorithms are generally
able to locate a local (and sometimes global) maximum of this 
function even in a space of large dimensionality. Similarly, the
covariance matrix of the errors can be found straightforwardly by
first numerically evaluating the Hessian matrix 
$\nabla\nabla\overline{\Pr}(\btheta|\vect{D})$ at the peak 
$\btheta=\hat{\btheta}$, and then calculating (minus) its inverse.
As we will see see in section~\ref{MCMCsampling}, however, this 
traditional approach to Bayesian parameter estimation 
has recently been superceded by Markov-Chain Monte-Carlo (MCMC)
techniques which allow one to explore the full posterior distribution,
without having to evaluate it over some large hypercube in parameter
space.

\subsection{Bayesian evidence and model selection}
\label{evidmodel}

Although the evidence term is usually ignored in the 
process of parameter estimation, it is central to selecting 
between different models for the data (see, for example, Sivia 1996). 
For illustration, let us
suppose we have two alternative models (or hypotheses) for 
the data $\vect{D}$; these hypotheses are 
traditionally denoted by $H_0$ and $H_1$. Let us assume further that
the model $H_0$ is characterised by the parameter set $\btheta$,
whereas $H_1$ is described by the set of parameters $\bphi$. 
For the model $H_0$, the probability density for an observed 
data vector $\vect{D}$ is given by
\begin{equation}
\Pr(\vect{D}|H_0) 
=  \int \Pr(\vect{D}|\btheta) \Pr(\btheta)\,d\btheta,
\label{eviddef0}
\end{equation}
where, on the left-hand side, we have made explicit the 
conditioning on $H_0$. Similarly, for the model $H_1$,
\begin{equation}
\Pr(\vect{D}|H_1) 
= \int \Pr(\vect{D}|\bphi) \Pr(\bphi)\,d\bphi.
\label{eviddef1}
\end{equation}
In either case, we see that the evidence is given by the average of
the likelihood function with respect to the prior. Thus, a model will
have a larger evidence if more of its allowed parameter space
is likely, given the data. Conversely, a model will have a small
evidence if there exist large areas of the allowed parameter space 
with low values of the likelihood, even if the likelihood function is strongly
peaked and the corresponding model predictions 
agree closely with the data. 
Hence the value of the evidence naturally incorporates
the spirit of Ockham's razor: a simpler theory, having a more
compact parameter space, will generally have a larger evidence than
a more complicated theory, unless the latter is 
significantly better at explaining the data.
The question of which of the models $H_0$ and $H_1$ is prefered 
is thus answered simply by comparing the relative
values of the evidences 
$\Pr(\vect{D}|H_0)$ and $\Pr(\vect{D}|H_1)$. The hypothesis
having the larger evidence is the one that should be
accepted. 

Unfortunately, the evaluation of an evidence integral, such as 
(\ref{eviddef0}), is a challenging numerical task. From (\ref{pbardef}), we
see that 
\begin{equation}
\Pr(\vect{D}|H_0) = \int\overline{\Pr}(\btheta|\vect{D})\,d\btheta,
\label{evid2}
\end{equation}
and so the evidence may only be evaluated directly if one can calculate
$\overline{\Pr}(\btheta|\vect{D})$ over some hypercube in parameter
space, which we noted earlier is often computational unfeasible. 
It is therefore common practice to approximate the posterior by
a Gaussian near its peak $\hat{\btheta}$, in which case it is
straightforward to show (see, for example, Hobson, Lahav \& Bridle 2002)
that an approximate expression for the evidence is
\begin{equation}
\Pr(\vect{D}|H_0) \approx (2\pi)^{M/2} |\vect{C}|^{1/2} \,
\Pr(\hat{\btheta})\Pr(\vect{D}|\hat{\btheta}),
\label{evidapprox}
\end{equation}
where $M$ is the number of parameters of interest $\btheta$ and
$\vect{C}$ is the estimated covariance matrix 
given in (\ref{covmatdef}). 
We note that, for (\ref{evidapprox}) to hold, the prior
and likelihood must be {\em correctly normalised}, such that
$\int\Pr(\btheta)\,d\btheta=1$ and
$\int\Pr(\vect{D}|\btheta)\,d\vect{D}=1$.

In a similar way to parameter estimation, however, this method
of approximating evidence values has recently been superceded by
MCMC techniques. As discussed below, by sampling from the posterior, 
is possible to calculate evidence values straightforwardly, without
recourse to evaluating the posterior over some large hypercube
in parameter space.

\section{Markov-Chain Monte-Carlo sampling}
\label{MCMCsampling}

As we have already commented, the traditional approaches to Bayesian parameter
estimation and evidence approximation outlined above have
recently become obsolete, at least for some types of
problems. Owing to the advent of faster computers and 
efficient algorithms, it has recently become numerically feasible to
sample directly from an (unnormalised) posterior distribution 
$\overline{\Pr}(\btheta|\vect{D})$ of large dimensionality
using {\em Markov-Chain Monte-Carlo} (MCMC) techniques. Indeed, 
in an astrophysical context, the 
MCMC approach has recently been applied to the determination of 
cosmological parameters from estimates of the cosmic microwave
background power spectrum (Christensen et al. 2001; Knox,
Christensen \& Skordis 2001).

The principles underlying MCMC sampling from some
posterior distribution have been discussed extensively elsewhere
(see, for example, Gilks, Richardson \& Spiegelhalter 1995), so we
shall give only a brief summary of the basic points.

\subsection{Sampling from the posterior}

Since it is numerically unfeasible to sample the
(unnormalised) posterior $\overline{\Pr}(\btheta|\vect{D})$
at a set of regular points (e.g. over some hypercube) in a parameter
space of large dimensionality, a natural alternative approach
is instead to {\em sample} from the posterior distribution. 

The major advantage to using a sampling-based approach is that,
once we have a set of samples $\{\btheta_1,\btheta_2,\ldots,\btheta_{N_{\rm
s}}\}$ from the posterior, we may use them to estimate the posterior
mean, mode or median, each of which can equally-well serve as
our `best' estimate $\hat{\btheta}$, depending on the problem under
consideration. Moreover, provided the posterior has been
sampled effectively, the mode should correspond to the
{\em global} maximum, and the presence of multiple peaks in the
posterior is readily observed. Using the samples,
it is also trivial to perform marginalisations over any subset of
the parameters $\btheta$. In particular, one may easily obtain the
one-dimensional marginalised (unnormalised) 
posterior distributions on each parameter $\theta_i$ separately, which
are given by
\[
\overline{\Pr}(\theta_i|\vect{D}) 
= \int \overline{\Pr}(\btheta|\vect{D})\,d\check{\btheta},
\]
where $d\check{\btheta} = d\theta_1\cdots d\check{\theta}_i\cdots 
d\theta_M$ denotes
that the integration is performed over all other parameters $\theta_j$ 
$(j \neq i)$. These can then be used to place confidence limits
on each parameter $\theta_i$ $(i=1,\ldots,M)$. Alternatively, the
samples may be used to determine the correlations between the 
estimates of different parameters.

\subsection{The MCMC method}

Sampling-based methods clearly offer enormous advantages
in Bayesian parameter estimation, but the difficult question remains
of how one efficiently obtains a set of samples 
$\{\btheta_1,\btheta_2,\ldots,\btheta_{N_{\rm s}}\}$ from a given
(unnormalised) posterior $\overline{\Pr}(\btheta|\vect{D})$.
Creating a set of independent samples from the posterior 
is very time consuming, and so it has become common practice to
use the MCMC approach, in which a Markov chain is constructed
whose equilibrium distribution is the required posterior.
Thus, after propagating the Markov chain for a given {\em burn-in}
period, one obtains (correlated) samples from the limiting
distribution, provided the Markov chain has converged.

A Markov chain is characterised by the fact that the state $\btheta_{n+1}$ 
is drawn from a distribution (or {\em transition kernel})
$\Pr(\btheta_{n+1}|\btheta_n)$ that
depends only on the previous state of the chain $\btheta_n$, and not
on any earlier state. The transition kernel is usually assumed not to
depend on $n$, in which case the Markov chain is {\em homogeneous}.
The standard approach to constructing a homogeneous Markov chain with a given
equilibrium distribution $p(\btheta)$ is to use the 
surprisingly simple {\em Metropolis-Hastings} (MH) algorithm, which is
based on the familiar notion of {\em rejection sampling}.
At each step $n$ in the chain, the next state $\btheta_{n+1}$ is chosen
by first sampling a {\em candidate} point $\btheta'$ from
some {\em proposal distribution} $q(\btheta|\btheta_n)$, which may in
general depend on the current state of the chain $\btheta_n$. The
candidate point $\btheta'$ is then accepted with probability
$\alpha(\btheta',\btheta_n)$ given by
\begin{equation}
\alpha(\btheta',\btheta_n)=\mbox{min}
\left[
1,\frac{p(\btheta')q(\btheta_n|\btheta')}{p(\btheta_n)q(\btheta'|\btheta_n)}
\right].
\label{mhalg}
\end{equation}
If the candidate point is accepted, the next state becomes 
$\btheta_{n+1}=\btheta'$, but if the candidate is rejected, the chain
does not move, so $\btheta_{n+1}=\btheta_n$. Remarkably, it is
straightforward to show that $q(\btheta|\btheta_n)$ can have {\em any}
form and the stationary distribution of the chain will be
$p(\btheta)$ (see, for example, Gilks et al. 1995).

Although, in theory, the convergence of the chain to the required
stationary distribution
is independent of choice of proposal distribution,
this choice is crucial in determining
both the efficiency of the MH algorithm and
the rate of convergence to the stationary distribution, as is seen
immediately by inspection of (\ref{mhalg}). The choice 
depends, in general, on the application under
consideration and on the form of the required stationary distribution.
Nevertheless, a discussion of some general principles
for choosing a proposal distribution is given by Kalos \& Whitlock (1986). 
Certain general classes of proposal distribution are commonly employed and 
the resulting special case of the MH algorithm often bears a
different name. For example, in the original 
{\em Metropolis} algorithm, the proposal distribution
is symmetric, such that $q(\btheta|\btheta_n)=q(\btheta_n|\btheta)$ for all
$\btheta$ and $\btheta_n$, and (\ref{mhalg}) simplifies accordingly. 
A special case of
the Metropolis algorithm is the {\em random-walk Metropolis}
algorithm, for which
$q(\btheta|\btheta_n)=q(|\btheta-\btheta_n|)$. Finally, an {\em
independence sampler} is a special case of the MH algorithm for which
$q(\btheta|\btheta_n)=q(\btheta)$, so that the proposal distribution
does not depend on the current state of the chain $\btheta_n$. Indeed,
Christensen et al. (2001) used an independence
sampler in which $q(\btheta)$ was taken simply to be the uniform
distribution over the parameter space, whereas Knox et al. (2001)
used a multivariate Gaussian as their proposal distribution.

In practice, the basic MH algorithm (or variations thereon) can be augmented
by the introduction of numerous speed-ups that allow the stationary
distribution to be sampled more efficiently, while still preserving
detailed balance. For example, so-called dynamical sampling methods
use gradient information about the posterior (see, for example, 
\'O'Ruanaidh \& Fitzgerald 1996).
We note, however, that gradient information is not available if
(some of) the parameters $\btheta$ are discrete, as will be the case
in the application of MCMC techniques to discrete object detection
in section~\ref{bayesobj}. Nevertheless, methods do exist for increasing the
efficiency of sampling without relying on gradient information (see,
for example, Skilling 2002).

\subsection{Burn-in, thermodynamic integration and the evidence}
\label{burnin}

As mentioned above, the states of the chain $\btheta_n$ 
can be regarded as samples from the stationary distribution
$p(\btheta)$ only after some initial {\em burn-in} period required for
the chain to reach equilibrium. Unfortunately, there exists no
formula for determining the length of the burn-in period, or for
confirming that a chain has reached equilibrium. Indeed, the topic
of convergence is still a matter of ongoing statistical research.
Nevertheless, several {\em convergence diagnostics} for determining
the length of burn-in have been proposed. These employ a variety of
theoretical methods and approximations that make use of the output
samples from the Markov chain. A review of such diagnostics is given
by Cowles \& Carlin (1994). It is worth noting, however, that
running several parallel chains, rather than a single long chain, 
can aid the diagnosis of convergence. Moreover, after burn-in,
the use of several parallel chains
can also increase the efficiency of sampling from a complicated
multimodal stationary distribution.

So far, we have not considered how MCMC sampling can be used to
evaluate the Bayesian evidence defined in (\ref{eviddef0}), in order to select
between different models for the data. A
straightforward approach to evidence evaluation using MCMC is provided
by the technique of {\em thermodynamic integration}
(see, for example, \'O'Ruanaidh \& Fitzgerald 1996).
Indeed, this
approach also provides a natural way of determining the length of the
burn-in period. Let us begin by defining
the quantity
\begin{equation}
E(\lambda) = \int [\Pr(\vect{D}|\btheta)]^\lambda\Pr(\btheta)\,d\btheta,
\label{thermo1}
\end{equation}
where we have raised the likelihood to the power $\lambda$.
From (\ref{eviddef0}), we see that the value of the evidence is given by
$E(1)$. Now suppose that, during the burn-in period, one begins
sampling from this modified posterior with $\lambda=0$ and then 
slowly raises the value according to some
{\em annealing schedule} until $\lambda=1$. This allows
the chain to sample from remote regions of the posterior
distribution. Indeed, by adopting an annealing schedule
based on the output of convergence diagnostics,
one can arrange for the end of the burn-in period to coincide with
the point at which $\lambda$ reaches unity.

During the burn-in period, one can use the Markov 
chain samples corresponding to a given value of $\lambda$
to obtain an estimate of the quantity
\begin{equation}
\langle \ln L \rangle_\lambda \equiv 
\frac{\int (\ln L) L^\lambda \Pr(\btheta) \,d\btheta}
{\int L^\lambda \Pr(\btheta) \,d\btheta},
\label{thermo2}
\end{equation}
where, for brevity, we have written $L=\Pr(\vect{D}|\btheta)$
for the likelihood. 
Comparing (\ref{thermo1}) and (\ref{thermo2}), we see that
\[
\langle \ln L \rangle_\lambda = \frac{1}{E}\nd{E}{\lambda}
=\nd{\ln E}{\lambda}.
\]
Thus, the (logarithm of) the evidence is given by
\[
\ln E(1) = \ln E(0) + \int_0^1 \nd{\ln E}{\lambda}\,d\lambda 
= \int_0^1 \langle\ln L\rangle_\lambda\,d\lambda,
\]
where we have used the fact that $E(0)=1$. Hence one may use the samples
obtained during the annealing period to obtain an estimate of the
evidence.

\subsection{The {\sc Bayesys} sampler}

In this paper, we use the implementation of the MCMC technique
in the {\sc Bayesys} software. This sampler uses the
Metropolis-Hastings algorithm, but coupled with a number of other
techniques that increase the efficiency with which the
stationary distribution is sampled, while maintaining detailed
balance. The sampler does not, however, make use of gradient information, so 
that discrete parameters can be easily accommodated. Evidence values
are calculated using the thermodynamic integration technique discussed
above. Multiple chains are also supported. A detailed discussion of 
the {\sc Bayesys} sampler is given by Skilling (2002).

\section{Bayesian object detection}
\label{bayesobj}

We now consider how the MCMC approach to Bayesian inference may be
used to address the difficult problem of detecting and characterising
discrete objects hidden in some background.
In order to keep our discussion as general as possible, let us
denote the totality of our available data by the vector
$\vect{D}$. This may represent the pixel values in a
single `image' (of arbitrary dimensionality) or collection of 
images, such as a multifrequency dataset. Equally, $\vect{D}$
could represent the Fourier coefficients of the image(s), or coefficients
in some other basis. In short, the exact specification of $\vect{D}$ 
is unimportant. We first consider the contribution to these data of
the discrete objects of interest.

\subsection{Discrete objects in a background}

Let us suppose we are interested in detecting and characterising
some set of (two-dimensional) 
discrete objects, each of which is described by a template
$\tau(\vect{x};\vect{a})$, which is parametrised in terms of
a set of parameters $\vect{a}$ that might typically 
denote (collectively) the position $(X,Y)$ of the object, its
amplitude $A$ and some measure $R$ of its spatial extent. For example,
circularly-symmetric Gaussian-shaped objects would
by defined by
\begin{equation}
\tau(\vect{x};\vect{a})=A\exp
\left[-\frac{(x-X)^2+(y-Y)^2}{2R^2}\right],
\label{objdef}
\end{equation}
so that $\vect{a} = \{X,Y,A,R\}$. If there exist $N_{\rm obj}$ such
objects in the dataset, we may write generically
\begin{equation}
\vect{D} = \vect{n}+ 
\vect{s}(\vect{a}_1,\vect{a}_2,\ldots,\vect{a}_{N_{\rm obj}}),
\label{splusn}
\end{equation}
where the `signal' vector $\vect{s}$ denotes the contribution to the data 
from the $N_{\rm obj}$ discrete objects, and $\vect{n}$ denotes the generalised
`noise' contribution to the data from other astrophysical emission
and instrumental noise. Although not a necessary requirement, 
in most applications the
contribution of the objects to the data is additive, in which
(\ref{splusn}) simplifies to
\[
\vect{D} = \vect{n}+ 
\sum_{k=1}^{N_{\rm obj}} \vect{s}(\vect{a}_k),
\]
where $\vect{s}(\vect{a}_k)$  denotes the contribution to the data 
from the $k$th discrete object. For simplicity we shall denote the
unknown parameters $N_{\rm obj}$ and 
$\vect{a}_k$ $(k=1,\ldots,N_{\rm obj})$ by the single
parameter vector $\btheta$. Clearly, we wish to use the data
$\vect{D}$ to place constraints on the values of the parameters
$\btheta$. 

\subsection{Defining the posterior distribution}

For any given parameterisation of the object template $\tau$, and
model of the background `noise' $\vect{n}$, one can write down
the likelihood function $\Pr(\vect{D}|\btheta)$. Additionally, one
may impose a prior $\Pr(\btheta)$ on the parameters. 
As discussed in section~\ref{bayesinf},
the Bayesian inference of the parameter values is then given by the entire
(unnormalised) posterior distribution
\begin{equation}
\overline{\Pr}(\btheta|\vect{D}) \equiv
\Pr(\vect{D}|\btheta)\Pr(\btheta).
\label{postdef}
\end{equation}
The problem of object identification and characterisation may then be 
addressed by sampling from this posterior using the MCMC techniques
described above.

As an example, suppose the 
data vector $\vect{D}$ contains the pixel values in a single
two-dimensional astronomical image, in which
the generalised background `noise' $\vect{n}$ corresponds to a 
statistically homogeneous Gaussian random field with 
covariance matrix $\vect{N} = \langle \vect{n}\vect{n}^{\rm
t}\rangle$. In this case, the likelihood function takes the form
\begin{equation}
\Pr(\vect{D}|\btheta) = 
\frac{\exp
\left\{-\tfrac{1}{2}{[\vect{D}-\vect{s}(\vect{a})]}^{\rm t}\vect{N}^{-1}
[\vect{D}-\vect{s}(\vect{a})]\right\}}
{(2\pi)^{N_{\rm pix}/2}|\vect{N}|^{1/2}},
\label{likedef}
\end{equation}
where $\vect{a}$ denotes collectively the parameter set
$\{\vect{a}_1,\vect{a}_2,\ldots,\vect{a}_{N_{\rm obj}}\}$.

The prior $\Pr(\btheta)$ on the parameters is also straightforward
to determine. Indeed, for most applications,
it is natural to assume that the number of objects $N_{\rm obj}$
and the parameters $\vect{a}_k$ for each object are mutually
independent, so that
\begin{eqnarray}
\Pr(\btheta) 
& = & \Pr(N_{\rm obj})\Pr(\vect{a}) \nonumber \\
& = & \Pr(N_{\rm obj})\Pr(\vect{a}_1)\Pr(\vect{a}_2)\cdots
\Pr(\vect{a}_{N_{\rm obj}}).
\label{priordef}
\end{eqnarray}
As mentioned above, the parameters $\vect{a}_k$, which 
characterise the $k$th object, will typically consist of its position
$X_k$ and $Y_k$, amplitude $A_k$ and spatial extent $R_k$, and
the priors imposed on these parameters will generally depend on
the application.
For example, one might
impose uniform priors on $X_k$ and $Y_k$ within the borders
of the image, whereas the priors on $A_k$ and $R_k$ may be provided
by some physical model of the objects one wishes to detect.
Similarly, one may impose a
prior on the number of unknown objects $N_{\rm obj}$, which is clearly
a discrete parameter. For example, if the objects of interest
are not clustered on the sky and have a mean number density 
$\mu$ per image area, then one would set 
\begin{equation}
\Pr(N_{\rm obj}) = \frac{\mu^{N_{\rm obj}}}{e^{\mu}N_{\rm obj}!}.
\label{poisson}
\end{equation}
\begin{figure*}
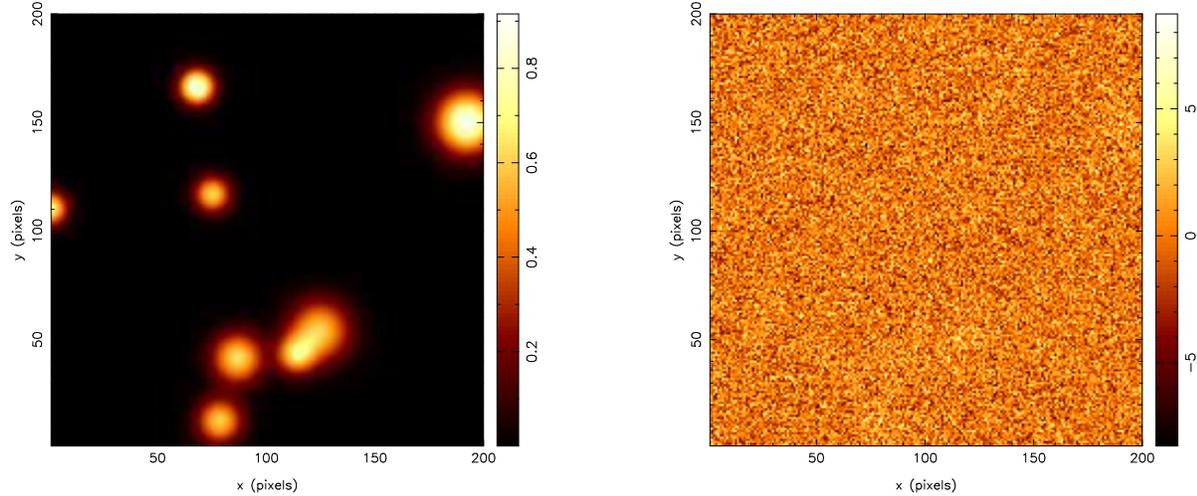

\begin{center}
\includegraphics[angle=-90,width=7cm]{fig1a.ps}
\qquad\qquad\qquad
\includegraphics[angle=-90,width=7cm]{fig1b.ps}
\caption{The toy problem discussed in section~\ref{toyprob}. The 
$200\times 200$ pixel test image (left panel)
contains 8 discrete Gaussian-shaped 
objects of varying widths and amplitudes;
the parameters $X_k$, $Y_k$, $A_k$ and $r_k$ for each object
are listed in Table~\ref{tab1}. The corresponding 
data map (right panel) has independent 
Gaussian pixel noise added with an rms of 2 units.}
\label{fig1}
\end{center}
\end{figure*}

It is clear from the above that a crucial complication inherent to the problem
of Bayesian object detection is that the length of the parameter
vector $\btheta$ is {\em variable}. In other words, the length
of $\btheta=\{N_{\rm obj},\vect{a}_1,\vect{a}_2,\ldots,
\vect{a}_{N_{\rm obj}}\}$ depends upon the value of $N_{\rm obj}$.
Thus, in the MH algorithm, the
proposal distribution must be able to propose moves between spaces of
differing dimension. In this case, the detailed balance conditions
must be carefully considered (Green 1994; Phillips \& Smith 1995).
The ability to sample from spaces of different dimensionality is
incorporated in the {\sc Bayesys} software.

\subsection{A toy problem}
\label{toyprob}

In order to illustrate the various approaches to Bayesian object
detection that we present below, we shall apply them to the simple
toy problem illustrated in Fig.~\ref{fig1}.
\begin{table}
\begin{center}
\begin{tabular}{crrrr}
\hline
Object & $X$  & $Y$ & $A$ & $R$ \\
\hline
1 &   0.7 & 110.5 & 0.71 & 5.34 \\
2 &  68.2 & 166.4 & 0.91 & 5.40 \\
3 &  75.3 & 117.0 & 0.62 & 5.66 \\
4 &  78.6 &  12.6 & 0.60 & 7.06 \\
5 &  86.8 &  41.6 & 0.63 & 8.02 \\
6 & 113.7 &  43.1 & 0.56 & 6.11 \\
7 & 124.5 &  54.2 & 0.60 & 9.61 \\
8 & 192.3 & 150.2 & 0.90 & 9.67 \\
\hline
\end{tabular}
\caption{The parameters $X_k$, $Y_k$ $A_k$ and $R_k$ $(k=1,\ldots,8)$
defining the Gaussian-shaped objects in Fig.~\ref{fig1}. The objects
are labelled in order of increasing $X$-value.}
\label{tab1}
\end{center}
\end{table}
The left panel shows our $200\times 200$-pixel test image, which
contains 8 Gaussian objects defined 
by (\ref{objdef}); the parameters $X_k$, $Y_k$ $A_k$ and $R_k$
$(k=1,\ldots,8)$ are listed in Table~\ref{tab1} in order of
increasing $X$-value. The $X$ and $Y$
position coordinates are drawn independently from the 
uniform distribution ${\cal U}(0,200)$ 
Similarly, the amplitude $A$ and size $R$ of each object
are drawn independently from the uniform distributions ${\cal U}(0.5,1)$ 
and ${\cal U}(5,10)$ respectively. In the right panel of
Fig~\ref{fig1}, we plot the corresponding data map, which has
independent (`white') Gaussian pixel noise added, with an rms of 2 units.
This corresponds to a signal-to-noise ratio of 0.25--0.5 as compared
to the peak emission in each object.
We see from the figure that, with this level of noise, 
no objects are visible to the naked eye,
and so this toy problem represents a considerable challenge for any
object detection algorithm.

\section{Simultaneous detection of all objects}
\label{simultaneous}

The theoretically most desirable approach is to attempt to detect and
characterise all the objects in the image simultaneously by
sampling from the (unnormalised) posterior distribution (\ref{postdef}),
with the likelihood $\Pr(\vect{D}|\btheta)$ given by (\ref{likedef}) and the 
prior $\Pr(\btheta)$ given by (\ref{priordef}). 
Thus, this approach allows one to include prior information
regarding the number of objects expected in the image. As mentioned
above, however, in this case the length of the parameter vector $\btheta$ is
variable, which can lead algorithmic complications. 
Moreover, if the
expected number of objects in the image is large, then so too 
will be the size of the corresponding parameter space that must be
sampled. As a result, the algorithm can be slow to burn-in and 
requires a large amount of CPU time. 

\begin{figure*}
\begin{center}
\includegraphics[width=6.1cm]{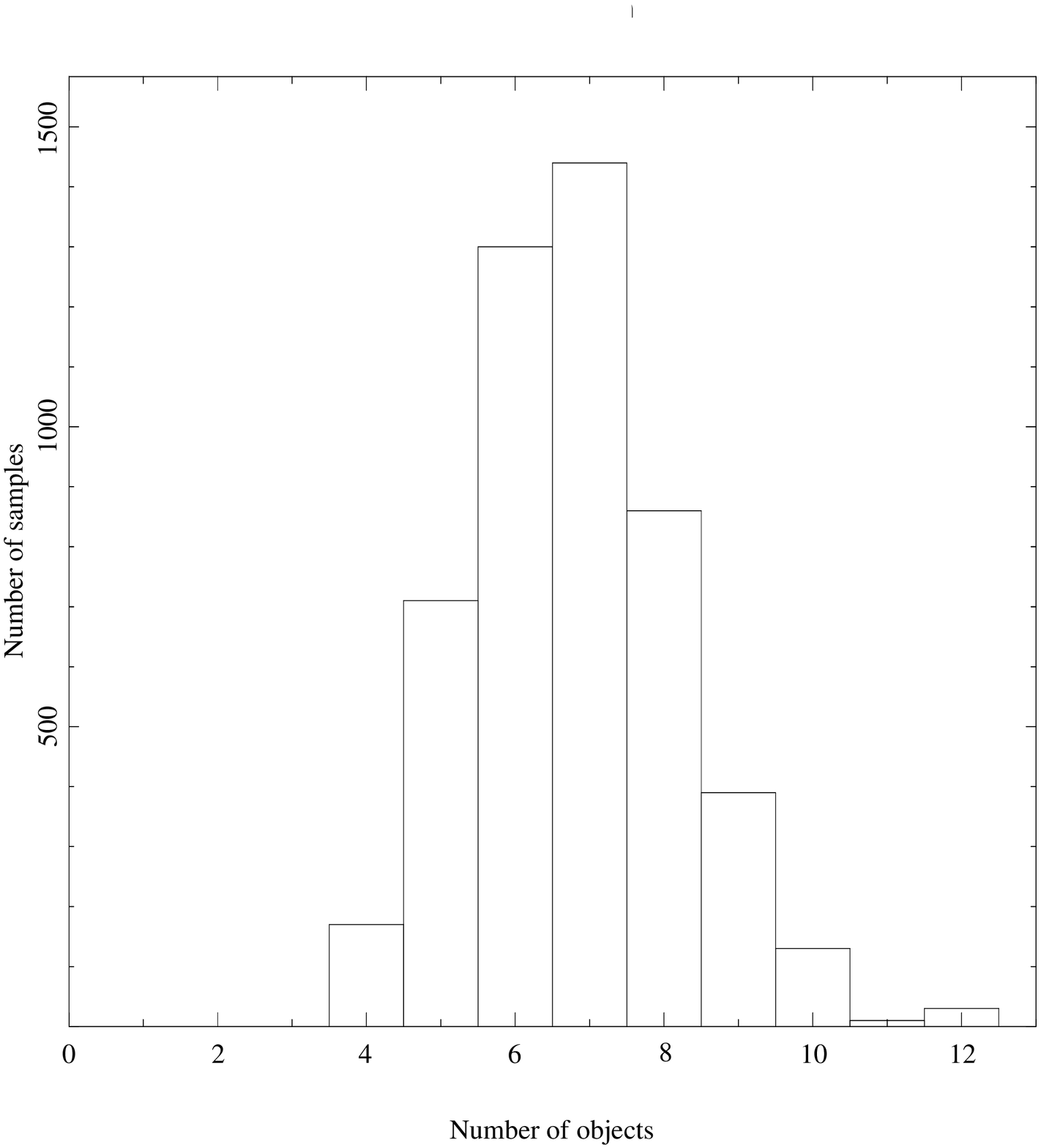}
\qquad\qquad\qquad
\includegraphics[width=6.4cm]{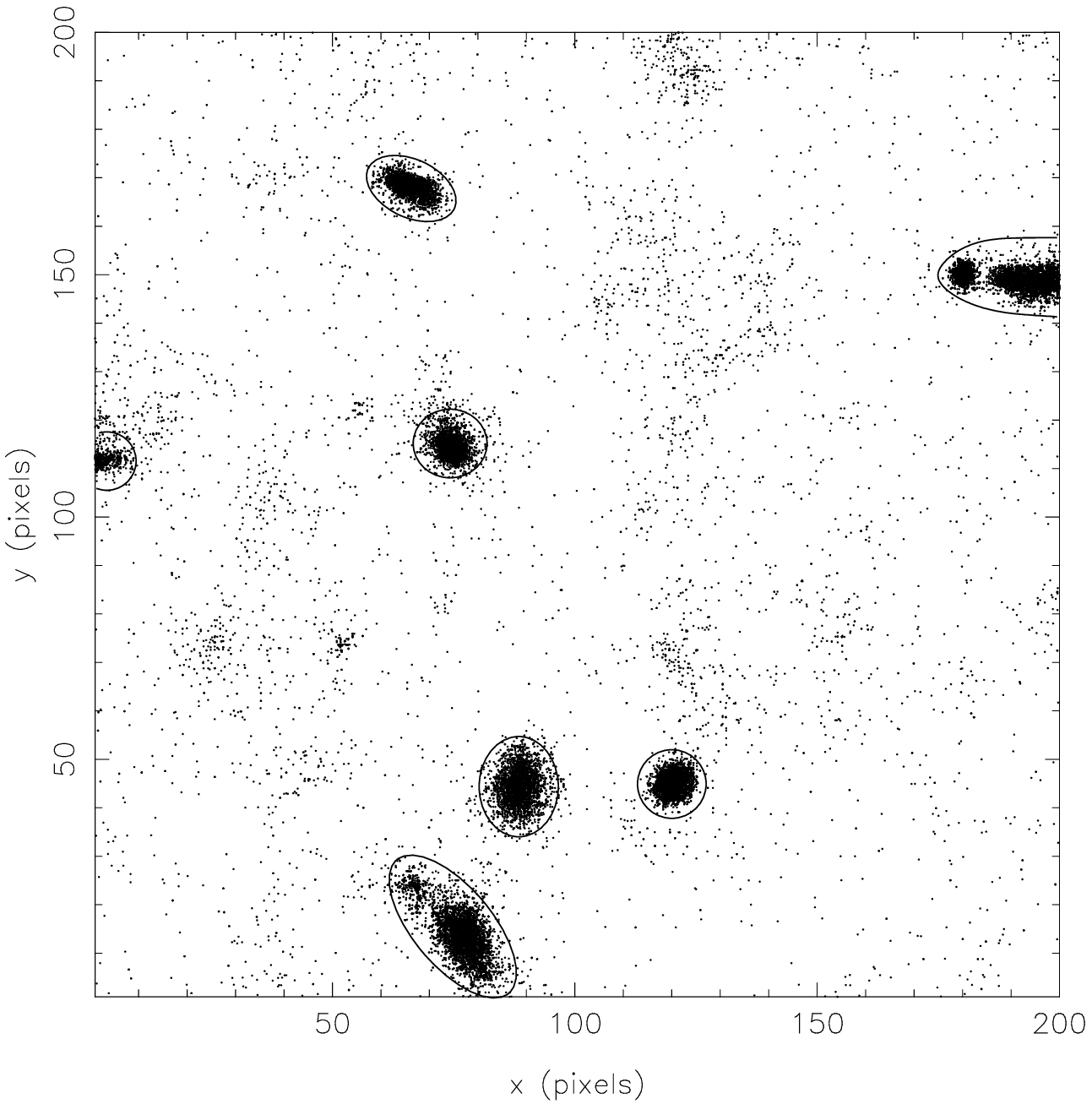}
\caption{Left: histogram of the number of post burn-in 
samples obtained in each subspace corresponding to a different number
of objects $N_{\rm obj}$. Right: the samples obtained for
$N_{\rm obs}=7$, projected into the $(X,Y)$-subspace; 
the ellipses indicate the samples used in the calculation
of the properties of the objects listed in Table.~\ref{tab2}.} 
\label{fig2}
\end{center}
\end{figure*}

In the analysis of the toy problem discussed above, we assume
the Poisson prior (\ref{poisson}) on the number of objects $N_{\rm obj}$,
with a mean of $\mu=4$, which is purposely chosen to be
somewhat smaller than the actual
number of objects $N_{\rm obj}=8$. Since the Poisson prior imposes
no upper limit on the possible number of objects, 
the overall parameter space under consideration
is formally the countably infinite union of subspaces
\[
\Theta  = \bigcup_{N_{\rm obj}=0}^\infty \Theta_{N_{\rm obj}},
\]
where $\Theta_{N_{\rm obj}} = \{\vect{a}_1,\ldots,\vect{a}_{N_{\rm obj}}\}$ 
denotes the $4N_{\rm obj}$-dimensional space corresponding
to the model with $N_{\rm obj}$ objects. The parameters of the $k$th object
are $\vect{a}_k=\{X_k,Y_k,A_k,R_k\}$, where we assume
(correctly) that $X_k$ and $Y_k$ $(k=1,\ldots,N_{\rm obj})$ are
drawn independently from the uniform distribution 
${\cal U}(0,200)$. For $A_k$ and $R_k$, we again (correctly) assume that
they are drawn independently from uniform distributions, but
we overestimate the width of these distribuion. In particular,
we assume $A_k$ and $R_k$ to be drawn from the uniform distributions 
${\cal U}(0,2)$ and ${\cal U}(3,12)$ respectively.

In sampling from the parameter space, we used 10 Markov chains and took
5000 post burn-in samples.
The results of this approach are illustrated in Fig.~\ref{fig2}.
In the left panel, we plot a histogram
of the number of samples obtained in each subspace of different
dimension, from which we note that the most favoured number of
objects is $N_{\rm obj} = 7$. 
One is free to use the 5000 post burn-in samples in a variety of
ways to place limits on the parameters $\btheta$. For illustration only,
in the right panel of Fig.~\ref{fig2}, we plot the samples obtained 
for the case $N_{\rm obs}=7$, projected into the $(X,Y)$-subspace.
We see that there exist seven main areas in which the samples are
concentrated, which we highlight with ellipses. Comparison with
Fig.~\ref{fig1} (left panel) shows that each of these areas
corresponds to a real object. The mean and standard 
deviation of the parameters $\{X_k,Y_k,A_k,R_k\}$ $(k=1,\ldots,7)$ 
for each detected object were calculated from
the samples contained in each ellipse. The results are given in
Table~\ref{tab2}, from which we see that the objects have been
characterised to reasonable accuracy. 
\begin{table}
\begin{center}
\begin{tabular}{crrrr}
\hline
Object & $X$  & $Y$ & $A$ & $R$ \\
\hline
1 &  $4.6\pm 3.3$  & $113.2\pm 3.4$ & $0.90\pm 0.52$ & $5.22 \pm 2.60$ \\
2 &  $65.8\pm 2.9$ & $168.2\pm 2.0$ & $0.97\pm 0.35$ & $4.70 \pm 1.27$ \\
3 &  $74.7\pm 2.5$ & $114.5\pm 2.8$ & $0.88\pm 0.42$ & $5.35 \pm 1.95$ \\
4 &  $76.4\pm 4.3$ & $14.2\pm 5.0$  & $0.69\pm 0.21$ & $9.28 \pm 2.14$ \\
5 &  $88.6\pm 2.9$ & $44.4\pm 3.7$  & $0.63\pm 0.23$ & $7.20 \pm 1.76$\\
6 and 7 &  $120.3\pm 1.9$ & $44.9\pm 1.9$ & $0.93 \pm 0.18$ & $9.45\pm 1.25$ \\
8 &  $188.7\pm 7.0$ & $149.2\pm 1.7$ & $1.22\pm 0.34$ & $6.76\pm 2.41$ \\
\hline
\end{tabular}
\caption{The mean and standard deviation of the
parameters $X_k$, $Y_k$ $A_k$ and $R_k$ $(k=1,\ldots,7)$
calculated from the samples contained within each ellipse in
Fig.~\ref{fig2}.}
\label{tab2}
\end{center}
\end{table}

We must note, however, that the two overlapping objects 6 and 7 in
Fig.~\ref{fig1} have been confused, and yield a single detected `object'. 
Moreover, for `objects'
4 and 8, the samples in Fig.~\ref{fig2} (right panel) 
are not tightly concentrated into a single cluster at 
the true position of the object.
Indeed, for `object' 8, there is some indication
that the samples are concentrated into two distinct regions, and
may in fact represent two distinct objects, one of which
is spurious. 

Clearly there exist more optimal strategies for using the samples to
characterise the objects and distinguish between real and spurious
detections. Although we address these issues in 
the next section, in the context of iterative object detection,
we shall not investigate further here.
The reason for this is the 
rather computationally-intensive nature of the above approach.
Using 10 chains and collecting 5000 samples after burn-in
required $\sim 5 \times 10^6$ evaluations of the posterior
distribution. Since the noise is uncorrelated the noise covariance
matrix $\vect{N}$ in the likelihood function (\ref{likedef}) is diagonal, and
so the posterior may be evaluated quickly. In 1 min of
CPU time on an Intel Pentium III 1 GHz 
processor, the posterior distribution can be evaluated $\sim 5000$
times. Nonetheless, the total analysis required $\sim 17$ hrs of CPU time.
As we discuss below, the process of object detection can be greatly
simplified by adopting an iterative procedure, and so we
shall not pursue the simultaneous detection algorithm further here.

\section{Iterative object detection}
\label{iterative}

An alternative approach to that discussed above, which
is both algorithmically simpler and considerably 
computationally faster, is to replace the prior (\ref{poisson}) by
\[
\Pr{(N_{\rm obj})}=\left\{
\begin{array}{ll}
1 & \mbox{if $N_{\rm obj}=1$} \\
0 & \mbox{otherwise}.
\end{array}
\right.
\]
In other words, our model for the data consists of just a single
object and so the full parameter space under consideration is
simply $\vect{a}=\{X,Y,A,R\}$, which is `only' 4-dimensional.
Although we have fixed $N_{\rm obj}=1$,
it is important to understand that this does {\em not}
restrict us to detecting just a single object in the data map.
Indeed, by modelling the data in this way, 
we would expect the posterior distribution to possess numerous
local maxima in the 4-dimensional parameter space, each 
corresponding to the location in this space of one of the objects.

\begin{figure*}
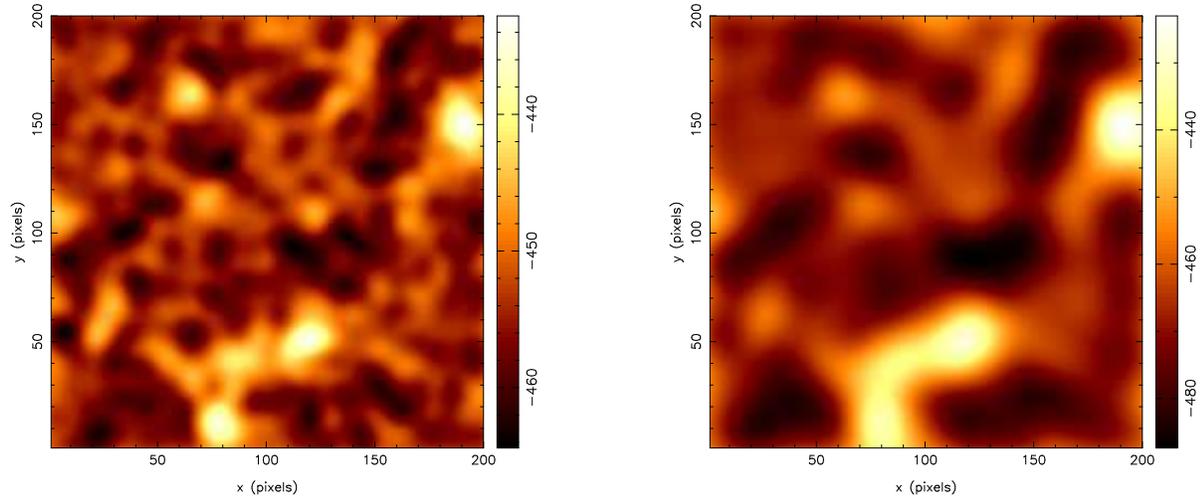

\begin{center}
\includegraphics[angle=-90,width=7cm]{fig3a.ps}
\qquad\qquad\qquad
\includegraphics[angle=-90,width=7cm]{fig3b.ps}
\caption{The 2-dimensional conditional log-posterior distributions 
in the $(X,Y)$-subspace for the toy problem illustrated in
Fig.~\ref{fig1}, where the model contains a single object
parametrised by $\vect{a}=\{X,Y,A,R\}$. The values of the 
amplitude $A$ and size $R$
are conditioned at $A=0.75$, $R=5$ (left panel) and
$A=0.75$, $R=10$ (right panel).}
\label{fig3}
\end{center}
\end{figure*}

This is best illustrated by direct evaluation of the posterior
distribution. As mentioned in section~\ref{bayesinf}, the ideal theoretical
solution to any Bayesian inference problem is to calculate the
full posterior distribution over some hypercube in parameter space
that contains all the probability.
However, even in the 4-dimensional parameter space we are considering 
here, this task is numerically intensive (indeed, this was the
original motivation for using MCMC sampling techniques). Nevertheless,
for the specific example at hand, 
the likelihood function (\ref{likedef}) has a
particularly simple form. It is thus computationally reasonable to
calculate the full posterior distribution over some {\em subspace}
of the 4-dimensional parameter space. It is most illustrative
to calculate the posterior distribution in the 2-dimensional
subspace defined by object position $(X,Y)$, while conditioning
on particular values of $A$ and $R$.

In Fig.~\ref{fig3}, we plot the 2-dimensional conditional log-posterior
distribution in the $(X,Y)$-subspace for $A=0.75$, $R=5$ (left panel)
and for $A=0.75$, $R=10$.  The value $A$ is chosen to be the mean of
the uniform distribution ${\cal U}(0.5,1)$ from which the amplitudes
of the objects were drawn, whereas the two values of $R$ correspond to
the limits of the uniform distribution ${\cal U}(5,10)$ from which the
sizes of the objects were drawn. Each conditional log-posterior
distribution is calculated on a $200\times 200$ grid in the
$(X,Y)$-subspace, which requires around 10 mins of CPU time on
an Intel Pentium III 1 GHz processor. We note that to calculate the full
4-dimensional log-posterior distribution at 200 points in each
direction would require $200^2 \times 10\mbox{ mins} \approx
280$ days of equivalent CPU time. 

We see from Fig.~\ref{fig3} that the conditional log-posterior
distributions contain multiple maxima and minima. As one might expect,
maxima do occur at the positions corresponding to each of the 8 objects
shown in Fig.~\ref{fig1}. We also note, however, that the
distributions contains numerous maxima that do {\em not} coincide with the
position of a real object, but instead occur because the background
noise in some areas has `conspired' to give the impression that an
object might be present. Unsurprisingly, this is particularly 
pronounced in the case $R=5$ (left panel). The effect is also easily
seen in the $R=10$ case (right panel), but the distribution is
correspondingly smoother, as one might expect. In either case, we see
that pronounced peaks in the log-posterior occur only for objects
2, 4, 7 and 8 (as listed in Table~\ref{tab1}). The peaks associated with
the remaining objects are not distinguishable by eye from `false' peaks
in the log-posterior that occur at positions where no object is present.
Finally, we note that for larger/smaller values of $A$ 
in the range $[0.5,1]$, the relative
height of the peaks in the posterior distribution at positions of true
objects increases/decreases slightly, but the overall shape of the
distribution remains very similar.

\subsection{Sampling of the posterior}

It is clear from Fig.~\ref{fig3} that the full 4-dimensional 
posterior distribution will be very complicated, possessing multiple
extrema. In particular, it is immediately obvious that any attempt
to detect objects by straightforward 
maximisation (e.g. gradient search) of the posterior 
distribution is doomed to failure. We therefore choose instead 
to sample from the posterior using the MCMC approach outlined in
section~\ref{MCMCsampling}. 

Several strategies present themselves for performing this sampling
of the posterior. The conceptually most straightforward approach
is to perform a `detailed' sampling of the full 4-dimensional posterior. 
This may be achieved in the following way. Firstly, the use of
several chains ($\sim$ the number of objects expected)
allows the sampler to explore full parameter space
more easily. Moreover, using a very slow annealing schedule
and a correspondingly long burn-in period during the thermodynamic
integration (see section~\ref{burnin}), affords the chains greater 
opportunity to sample remote regions of the posterior distribution. 
Finally, after burn-in, a large number of samples are taken.

In general, however, the use of multiple chains, a long-burn and a
large number of samples make this approach very time consuming, as was
the case for the simultaneous detection of objects discussed in the
previous section. This
is particularly true, when the posterior distribution is
dominated by a pronounced peak (or set of peaks) corresponding to one
(or more) object. This can occur, for example, if the true amplitudes $A$ of
some of the objects are much larger than the others, or simply by
chance in cases where the signal-to-noise ratio is somewhat higher
than that used in our toy problem. In this case, a significant fraction of
the samples obtained are in the neighbourhood(s) of the pronounced
peak(s), and so a large total number of samples are needed in order to
obtain a reasonable representation of the full posterior
distribution. We shall therefore not pursue
the `detailed' sampling approach here. 

\subsubsection{The {\sc McClean} algorithm}
\label{mccleansec}

The drawbacks associated with the above method do themselves, however, 
suggest an alternative {\em iterative} approach to the problem, in
which one attempts to detect and characterise one (or a few) object 
at a time. In this case, one is not concerned with `detailed' sampling
of the full posterior distribution. Instead, one is content with
sampling the distribution adequately in the neighbourhood of its most
pronounced peak(s). This can be performed efficiently using only
a few chains, a relatively fast annealing schedule during the
thermodynamic integration, and requires many fewer post burn-in
samples. Hence, this approach is significantly computationally faster
than the `detailed' sampling of the full posterior distribution. 
Having characterised the object(s) detected in this way, it (they)
can then be subtracted from the data and the process repeated.
In this way, the new log-posterior distribution will no longer
contain the pronounced peak(s) associated with the subtracted
object(s), but will instead be dominated by peaks associated with the
most significant remaining object(s). This procedure has some
features reminiscent of the widely-used {\sc Clean} 
algorithm (H\"ogbom 1974) for
producing astronomical images from radio-inteferometer data. 
Since our approach is based on MCMC sampling techniques, we thus call it
the {\sc McClean} algorithm.

One remaining question is when to stop the iterative process. In fact,
this may be answered straightforwardly using the notion of Bayesian
evidence, discussed in section~\ref{evidmodel}. Let us denote the model
(hypothesis) that there are no objects (remaining) in the image by
$H_0$, and the model consisting of a single (remaining) object, i.e. $N_{\rm
obj}=1$, by $H_1$. Since $H_0$ has no parameters associated with it,
the evidence $\Pr(\vect{D}|H_0)$
is simply the value of the likelihood function (\ref{likedef}) evaluated
for the case with no objects in the data. The model $H_1$, however,
depends on the parameters $\vect{a} = \{X,Y,A,R\}$, and the evidence
is given by
\[
\Pr(\vect{D}|H_1) = \int
\Pr(\vect{D}|\vect{a})\Pr(\vect{a})\,d\vect{a}.
\]
As shown in section~\ref{evidmodel}, 
the value of the evidence $\Pr(D|H_1)$ may
be evaluated (to a good approximation) from the MCMC samples
using thermodynamic integration.  Thus, after each iteration of
the {\sc McClean} algorithm, one calculates the evidence ratio
$\Pr(\vect{D}|H_1)/\Pr(\vect{D}|H_0)$. If this ratio is greater than
unity one accepts the identified object(s) and repeats the
procedure. If the ratio is less than unity, the object(s) identified
in the last iteration are discarded, and the algorithm is stopped.

From its design, it is clear that
the {\sc McClean} algorithm is a compromise between the full Bayesian
approach outlined in section~\ref{simultaneous} and the 
desire to obtain results in a reasonable amount of CPU time.
For the toy problem at hand, it is clearly a sensible approach, which
we show below yields good results. As an approximation to the fully
Bayesian method, however, there will inevitably be situations in which
its performance is poorer. A discussion of straightforward refinements to the
basic {\sc McClean} algorithm, which broaden its range of applicability, 
while retaining its computational
efficiency, are given in section~\ref{conc}.

\subsubsection{Application to the toy problem}
\label{mctoy}

We now apply the {\sc McClean} algorithm to the toy problem
discussed in section~\ref{toyprob} and illustrated in Figs~\ref{fig1} and
\ref{fig3}. 
\begin{figure*}
\begin{center}
\includegraphics[angle=-90,width=14cm]{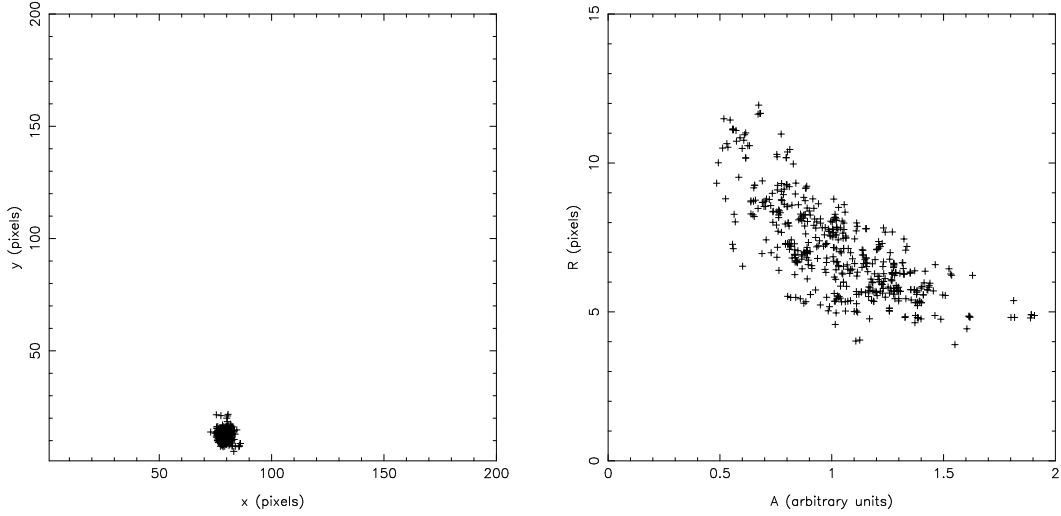}
\caption{The 500 samples from the posterior obtained on
the third iteration of the {\sc McClean} algorithm when
applied to the toy problem. The samples are projected into
the 2-dimensional subspaces $(X,Y)$ (left panel) and
$(A,R)$ (right panel).}
\label{fig4}
\end{center}
\end{figure*}
We perform the MCMC sampling using 5 chains and, at each iteration,
500 post burn-in samples
are taken from the posterior, and the algorithm
identifies 6 objects before stopping. The total analysis required
$\sim 2 \times 10^5$ evaluations of the posterior, which took
$\sim 40$ mins of CPU time on a single Intel Pentium III 1 GHz
processor; this is clearer considerably less computationally intensive
than attempting to detect all the objects simulaneous, as discussed
in section~\ref{simultaneous}.

The detected objects are listed in Table~\ref{tab3} in the order in
which they are identified. For each identified object, the
table also lists the mean
and standard deviation of the 1-dimensional marginalised distribution
for each parameter $\{X,Y,A,R\}$.
\begin{table}
\begin{center}
\begin{tabular}{crrrr}
\hline
Object & $X$\phantom{aaaa}  & $Y$\phantom{aaaa} & $A$\phantom{aaaa} & $R$
\phantom{aaaa} \\
\hline
8   &  $192.6\pm 2.5$ & $149.0\pm 2.0$ & $0.93\pm 0.17$ & $9.47\pm 1.33$ \\
6 and 7 &  $118.3\pm 2.3$ & $50.5\pm 1.9$  & $0.90\pm 0.24$ & $8.32\pm 1.65$ \\
4 &  $79.0\pm 1.9$ & $12.2\pm 2.2$ & $1.04\pm 0.26$ & $7.08\pm 1.57$ \\
5 &  $86.6\pm 3.2$ & $40.5\pm 2.5$ & $0.63\pm 0.19$ & $8.10\pm 1.61$ \\
2 &  $64.9\pm 2.4$ & $164.0\pm 2.3$ & $0.74\pm 0.32$ & $5.94\pm 1.51$ \\
1 &  $4.2\pm 2.9$ & $108.7\pm 3.1$ & $0.61\pm 0.27$ & $6.41\pm 1.92$ \\
\hline
\end{tabular}
\caption{The objects identified by the {\sc McClean} algorithm when
applied to the toy problem discussed in section~\ref{toyprob}. The
objects are listed in the order in which they were identified.
The mean and standard deviation of the 1-dimensional marginalised
distributions for each parameter $\{X,Y,A,R\}$
are listed for each identifed object.}
\label{tab3}
\end{center}
\end{table}
Comparing these values with those listed in Table~\ref{tab1}, we see that
the algorithm has accurately recovered the true parameter values
for objects 8,4,5,2 and 1, within the stated errors. The second
identified `object' listed in Table~\ref{tab3} is, however, a composite
of the two real overlapping objects 6 and 7, which the algorithm has
been unable to separate with the applied level of pixel noise.
The algorithm also fails to identify object 3. In fact, if the
algorithm is allowed to continue past the point where the evidence
criterion suggests termination, one finds that some samples from the
posterior are clustered in the neighbourhood of object 3. However,
samples in these later iterations are also clustered 
in other areas where no real objects are located,
but maxima nevertheless exist in the posterior (as illustrated in
Fig.~\ref{fig3}). Thus, the evidence-based stopping criterion provides
a robust method for avoiding spurious detections.

\begin{figure*}
\begin{center}
\includegraphics[angle=-90,width=14cm]{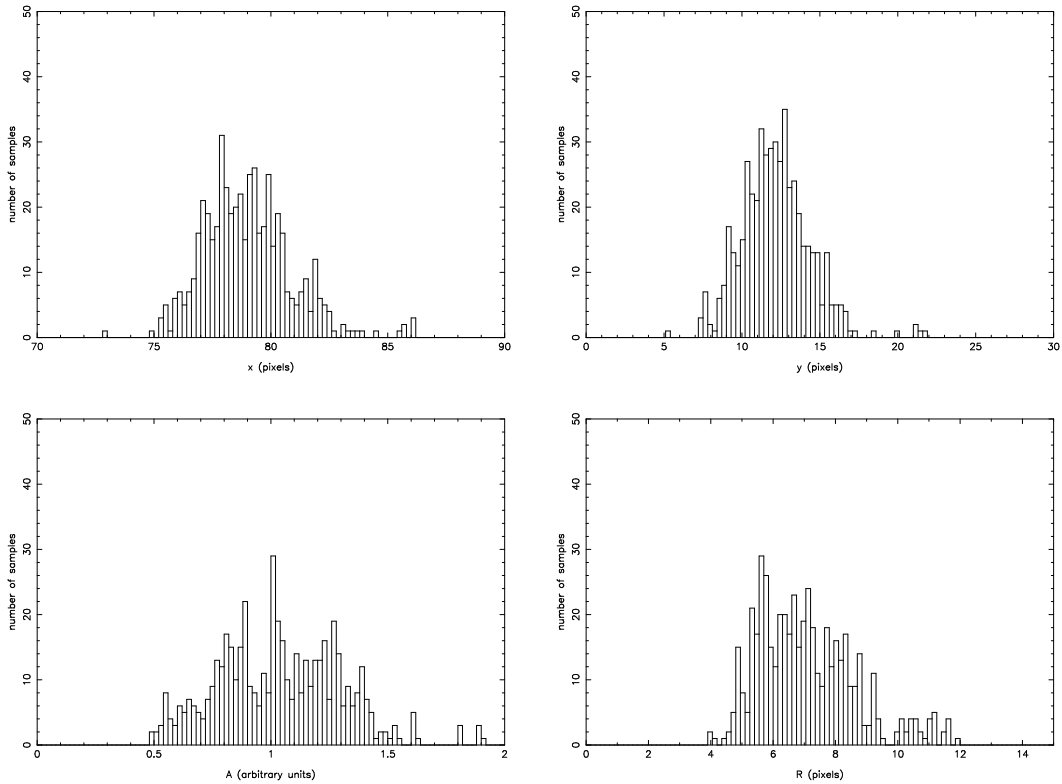}
\caption{The 500 samples from the posterior obtained on
the third iteration of the {\sc McClean} algorithm, projected into
1-dimensional subspaces to yield the marginalised
distributions for the parameters $X$, $Y$, $A$ and $R$ separately.}
\label{fig5}
\end{center}
\end{figure*}

For each identified object, the samples from the posterior clearly
contain more information
than simply the mean and standard deviation of its defining
parameters. As an illustration, in Fig.~\ref{fig4} we plot the 500 samples
obtained on the third iteration of the algorithm (which identified 
object 4) projected into the two 2-dimensional subspaces
$(X,Y)$ and $(A,R)$.
By projecting these samples further into 1-dimensional subspaces, we
obtain four marginalised distributions for the parameters
$X$, $Y$, $A$ and $R$ separately; these are shown in Fig.~\ref{fig5}.

The number of objects detected by the algorithm,
and the accuracy with which their
defining parameters are determined, will clearly
depend on the rms level of the added pixel noise. For comparison, 
in Table~\ref{tab4}, we list the objects detected
for the case where the rms
noise level is increased to 3 units. This noise level
corresponds to a signal-to-noise ratio of $0.16-0.33$ 
as compared with the peak values of the objects.
\begin{table}
\begin{center}
\begin{tabular}{crrrr}
\hline
Object & $X$\phantom{aaaa}  & $Y$\phantom{aaaa} & $A$\phantom{aaaa} & $R$
\phantom{aaaa} \\
\hline
4 &  $79.1\pm 1.8$ & $11.1\pm 2.6$ & $1.19\pm 0.30$ & $6.95\pm 1.33$ \\
8 &  $192.5\pm 3.1$ & $149.5\pm 2.6$ & $1.04\pm 0.31$ & $8.73\pm 1.95$ \\
6 and 7 &  $118.4\pm 2.5$ & $51.2\pm 2.0$  & $1.21\pm 0.39$ & $6.40\pm 1.90$ \\
5 &  $85.6\pm 7.1$ & $39.1\pm 5.6$ & $0.58\pm 0.32$ & $8.10\pm 2.46$ \\
\hline
\end{tabular}
\caption{As in Table~\ref{tab3}, but for an rms noise level of 3 units.}
\label{tab4}
\end{center}
\end{table}
We see that, in the presence of a noise background with a larger rms, the
algorithm detects only 4 distinct objects, with the real objects
6 and 7 again combined into a single detection. Nevertheless, the
defining parameters of those object identified are once again
accurate to within the derived errors.

It is also of interest to consider lower noise levels. In
Table~\ref{tab5}, we list the objects detected in the case for which the
rms noise level is decreased to 1 unit, which 
corresponds to a signal-to-noise ratio of $0.5-1$ 
as compared with the peak values of the objects. 
\begin{table}
\begin{center}
\begin{tabular}{crrrr}
\hline
Object & $X$\phantom{aaaa}  & $Y$\phantom{aaaa} & $A$\phantom{aaaa} & $R$
\phantom{aaaa} \\
\hline
8   &  $192.3\pm 1.3$ & $149.7\pm 1.0$ & $0.90\pm 0.08$ & $9.70\pm 0.83$ \\
6 and 7 &  $118.6\pm 1.3$ & $50.1\pm 1.1$  & $0.72\pm 0.09$ & $9.7\pm 0.89$ \\
4 &  $79.0\pm 1.0$ & $12.6\pm 1.2$ & $0.79\pm 0.13$ & $7.24\pm 0.99$ \\
5 &  $86.2\pm 1.3$ & $40.8\pm 1.2$ & $0.63\pm 0.10$ & $7.81\pm 0.79$ \\
2 &  $66.6\pm 1.0$ & $165.1\pm 0.85$ & $0.93\pm 0.19$ & $4.99\pm 0.74$ \\
3 &  $73.8\pm 1.5$ & $116.1\pm 1.5$ & $0.64\pm 0.15$ & $5.22\pm 0.79$ \\
1 &  $2.4\pm 1.1$ & $109.0\pm 1.57$ & $0.67\pm 0.18$ & $5.85\pm 1.32$ \\
\hline
\end{tabular}
\caption{As in Table~\ref{tab3}, but for an rms noise level of 
1 unit.}
\label{tab5}
\end{center}
\end{table}
In this case, we see that the algorithm detects all the
real objects with high accuracy, except for once again combining the
overlapping objects 6 and 7 into a single detection.
Indeed, this degeneracy can only be broken when the rms noise level
falls below $\sim 0.5$ units. This is easily understood if one
plots the conditional log-posterior in the $(X,Y)$-subspace
(i.e. analogous to Fig~\ref{fig3}) for some typical 
values of $A$ and $R$ and for various noise levels. Only for $\sigma \la
0.5$ units does the log-posterior distribution in the location of
objects 6 and 7 divide into two distinct maximia.
In this case, as one would expect, the algorithm 
identifies objects 6 and 7 as separate features, and finds the correct defining
parameters for each object within the derived errors. It is worth
noting in Tables~\ref{tab3} \& \ref{tab5} that the recovered $X$-position for 
object 1 is slightly less accurate than for the other objects, since
it is located at the edge of the observed field.

\subsection{Global maximisation of the posterior}

Instead of sampling from the posterior
using MCMC techniques, one can obtain a less
theoretically-desirable, but much faster, solution by simply locating
the {\em global} maximum of the posterior at each iteration.  The
price to be paid for dispensing with the sampling approach is that one
must return to the more traditional method, outlined in
section~\ref{bayesinf}, of characterising the posterior distribution in terms
of a set of best estimates
$\hat{\vect{a}}=\{\hat{X},\hat{Y},\hat{A},\hat{R}\}$ that locate its
global maximum.  Similarly, one must be content with describing the
uncertainties in the derived parameters in terms of the estimated
covariance matrix given by (minus) the inverse of the Hessian matrix
at the global maximum.  Nevertheless, this approach is perfectly adequate
for most applications. 

\subsubsection{The {\sc MaxClean} algorithm}

It is clear from the conditional log-posterior distributions 
for the toy problem, shown in Fig.~\ref{fig3}, that
a numerical maximiser (minimiser) which locates only a local maximum
will most often produce a spurious detection. One must
instead seek to locate the global maximum of the posterior. 
In fact, an MCMC sampler can itself be used to 
locate the global maximum in parameter space. This is most easily
achieved by introducing an annealing parameter
similar to that used in thermodynamic integration (see section
\ref{burnin}). In this case, however, one 
raises the full (unnormalised) posterior
$\overline{\Pr}(\btheta|\vect{D})$ to the power $\lambda$.
Allowing this parameter to continue increasing
gradually {\em beyond} unity artificially `sharpens' the posterior
distribution. As $\lambda$ grows the samples are then concentrated
into an increasingly compact region of parameter space, and eventually
locate the global maximum to some required accuracy. 
Nevertheless, even if one does not have access to a
reliable MCMC sampler, one may still perform Bayesian object detection
and characterisation using elementary and widely-available algorithms,
as we illustrate below.

The most widely-used technique for performing locating a global
extremum is
{\em simulated annealing}. In particular, we use the downhill simplex
implementation of a simulated annealing minimiser suggested by
Press et al. (1994). In short, this algorithm behaves in the same way
as a standard downhill simplex minimiser, except that 
one adds a positive, 
logarithmically-distributed random variable, proportional to the
{\em annealing temperature} $T$, to the function value associated with
each vertex of the simplex. Then a similar random variable is
subtracted from the function value of every new point that is tried as
a replacement vertex. In this way, the algorithm always accepts a
true downhill step, but sometimes accepts an uphill one. 
Indeed, for a finite $T$-value, the algorithm can wander
freely among local minima of depth less than about $T$. As $T$ is
reduced according to some {\em annealing schedule}, the number of
minima qualifying for frequent visits is reduced and, 
in the limit $T \to 0$, the algorithm reduces exactly to the standard
downhill simplex minimisation method and converges to the nearest
local minimum. If the annealing schedule reduces the temperature $T$
sufficiently slowly, it is very likely that the simplex will shrink
into the region containing the global minimum. 

Possible choices for the annealing
schedule are discussed by Press et al. (1994). We found that an effective
approach was simply to reduce $T$ from some initial starting value
$T_0$ by a fixed factor $f$, after every
$N_{\rm iter}$ moves of the simplex vertices. The choice of $T_0$,
$f$ and $N_{\rm iter}$ are problem-specific, and are discussed
below with reference to the toy problem.

Once one has a technique for reliably locating the global maximum of
the posterior distribution, the basic iterative procedure for 
object detection is analogous to the {\sc McClean} algorithm discussed
above. In this case, however, only one object can be identified
and removed in each iteration. We call the resulting procedure
the {\sc MaxClean} algorithm.

Finally, one must reconsider the
problem of when to stop the algorithm. Ideally, one would like to
use the evidence ratio $\Pr(\vect{D}|H_1)/\Pr(\vect{D}|H_0)$
as our stopping criterion,  as discussed above for the {\sc McClean}
algorithm. Once again $\Pr(\vect{D}|H_0)$ is simply the value
of the likelihood function evaluated for the model containing no
objects. In the absence of samples
from the posterior, however, one cannot obtain the evidence 
$\Pr(\vect{D}|H_1)$ for the single-object model by performing a thermodynamic
integration. Nevertheless, an approximate value for the evidence 
$\Pr(\vect{D}|H_1)$ after each iteration can
be obtained by approximaing the posterior distribution as a
4-dimensional multivariate Gaussian about the 
corresponding global maximum found by the simulated annealing simplex
algorithm. As discussed in section~\ref{evidmodel}, an approximate value for
the evidence is then given by (\ref{evidapprox}). The covariance matrix
$\vect{C}$ appearing in this expression is simply (minus) the inverse
of the Hessian matrix of the log-posterior at the peak. In turn, we find that
a good approximation to this Hessian matrix is obtained by
using simple second-differencing algorithm along each parameter
direction. Hence, the Gaussian approximation to the evidence 
$\Pr(\vect{D}|H_1)$ may be easily and quickly calculated, and the resulting
(approximate) evidence ratio can be used to determine when to halt the
algorithm.

\subsubsection{Application to the toy problem}

We now apply the {\sc MaxClean} algorithm to the toy problem
discussed in section~\ref{toyprob} and illustrated in Figs~\ref{fig1} 
\& \ref{fig3}, for the
case in which the rms of the pixel noise is 2 units. 
As mentioned above, in order to apply the {\sc MaxClean} algorithm one
must first define an effective annealing schedule. We find that a
robust approach is as follows: starting from some initial annealing
temperature $T_0$, after every $N_{\rm iter}$ moves of the
simplex vertices one reduces $T$ by a fixed factor $f$. The choice of
$T_0$, $f$ and $N_{\rm iter}$ clearly affects the total required number of
evaluations of the posterior distribution, and hence the overall speed
of the algorithm. This choice also affects the efficiency with which the
global maximum can be located. 

\begin{figure*}
\begin{center}
\includegraphics[width=15cm]{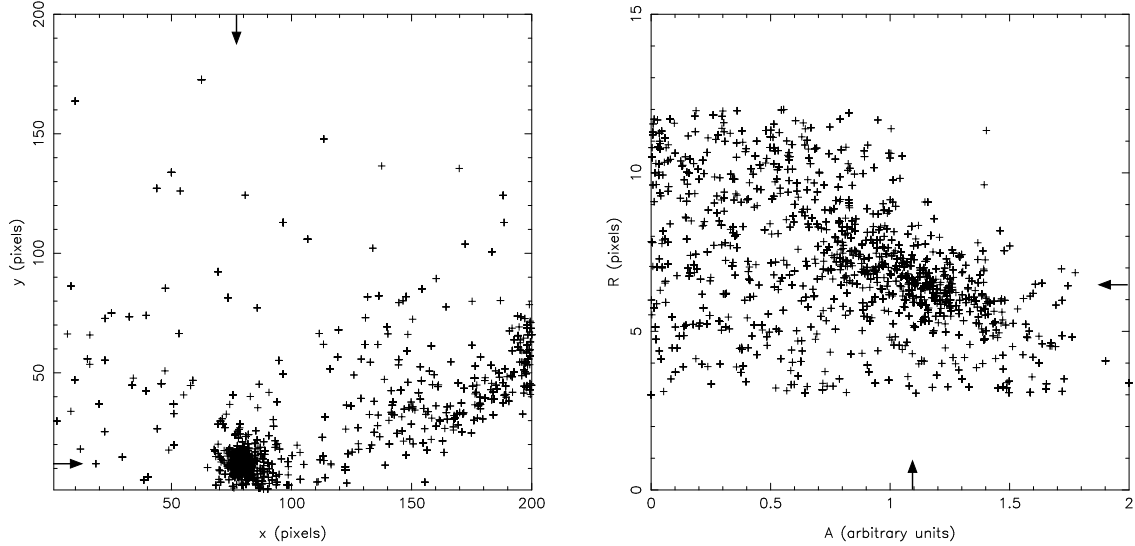}
\caption{The position of the `best-point' in the simplex after each
move, during the third iteration of the {\sc MaxClean} algorithm
(in which object 4 is detected). Each point is plotted simultaneously
in the $(X,Y)$ and $(A,R)$-subspaces. The position to which the
simplex converged is indicated by the short arrows.}
\label{fig6}
\end{center}
\end{figure*}

As stated above, for a finite value of
$T$, the simplex can wander freely among local maxima (minima) of
height less than about $T$. Thus one must set $T_0$ to be at least of the order
of the largest variations one expects in the log-posterior
distribution. From Fig.~\ref{fig3}, we see that
log-posterior varies by $\sim 70$ units in the $(X,Y)$-subspace
corresponding to the conditioned values $A=0.75$, $R=5$ and 
$A=0.75$, $R=10$. In order that the simplex has access to all regions
of the log-posterior in the early stages of the annealing schedule,
we thus set $T_0=100$. We note that, in the absence of this prior
information one could set $T_0$ to some arbitrarily large value,
but at the cost of requiring a larger total number of function
evaluations before the algorithm converges. In choosing the fixed factor
$f$, one must ensure that the annealing schedule reduces the
temperature `slowly enough' that the algorithm converges to the 
global maximum of the log-posterior. We find that the convergence of
the algorithm is reasonably insensitive to the precise value of $f$, provided
it lies in the range $1 \la f \la 4$; we choose $f=1.5$ for the
results presented below. Finally, the parameter $N_{\rm iter}$
determines the number of moves of the downhill simplex performed 
at each value of the annealing temperature $T$ (until convergence is
obtained). We again find the
convergence of the algorithm to the global maximum is reasonably
insensitive to this parameter, provided $N_{\rm iter} \ga 50$; we
choose $N_{\rm iter}=200$ for the results presented below.

We find that the {\sc MaxClean} algorithm works very well on the toy problem 
and produces results similar to those presented in section~\ref{mctoy},
obtained using {\sc McClean}. 
\begin{table}
\begin{center}
\begin{tabular}{crrrr}
\hline
Object & $X$\phantom{aaaa}  & $Y$\phantom{aaaa} & $A$\phantom{aaaa} & $R$
\phantom{aaaa} \\
\hline
8 &  $191.3\pm 2.5$ & $149.5\pm 2.1$ & $0.99\pm 0.10$ & $8.59\pm 0.69$ \\
6 and 7 &  $118.8\pm 1.9$ & $51.0\pm 1.8$ & $0.99\pm 0.24$ & $7.64\pm 0.67$ \\
4 &  $78.6\pm 1.5$ & $11.8\pm 1.7$ & $1.10\pm 0.17$ & $6.51\pm 0.98$ \\
5 &  $86.8\pm 3.6$ & $40.3\pm 2.5$  & $0.64\pm 0.12$ & $8.38\pm 1.40$ \\
2 &  $64.9\pm 2.3$ & $164.1\pm 2.0$ & $0.86\pm 0.34$ & $5.36\pm 2.62$ \\
1 &  $0.2 \pm 5.1$ & $108.3 \pm 2.8$ & $0.92\pm 0.24$ & $6.47 \pm 1.65$ \\
\hline
\end{tabular}
\caption{The objects identified by the {\sc MaxClean} algorithm when
applied to the toy problem discussed in section~\ref{toyprob}. The
objects are listed in the order in which they were identified.
The derived parameters correspond to the global maximum of the
posterior distribution at each iteration of the algorithm. The
quoted errors on the parameters of each object refer to the 
square-root of the corresponding diagonal element of the covariance
matrix derived from the Hessian at the peak.}
\label{tab6}
\end{center}
\end{table}
The objects identified are listed in Table~\ref{tab6}. We note that the
order in which they are detected by the {\sc MaxClean} algorithm
is the same as that produced by {\sc McClean}. Indeed, one would
expect this to be the case, since at each iteration each algorithm
either samples the posterior in the neighbourhood of the
global maximum or locates its position directly. We also see from
Table~\ref{tab6} that the
parameters defining the identified objects are in reasonable agreement
with the true values listed in Table~\ref{tab1} and are similar
to those obtained using the {\sc McClean} algorithm, listed in
Table~\ref{tab3}. It must be remembered, however, that, for each object,
the derived parameter values and estimated errors produced by the {\sc McClean}
and {\sc MaxClean} algorithms are defined in different ways.
In {\sc McClean} they correspond respectively to the mean and 68-per
cent confidence limits
of the 1-dimensional marginalised distribution for each parameter as
derived from the MCMC samples. In {\sc MaxClean}, however, the estimated
parameter values correspond simply to the location of the 
global maximum of the 
multi-dimensional posterior and the errors are approximated by
the square-root of the appropriate diagonal element of the covariance
matrix, derived from the Hessian at the peak. We also note in
Table~\ref{tab6} that the quoted error on the recovered $X$-position
of object 1 is larger that for the other objects, since it lies
at the edge of the observed field.

In Fig.~\ref{fig6}, we plot 
an illustration of the behaviour of the simulated-annealing
downhill simplex miminiser on the third iteration of the
{\sc MaxClean} algorithm (in which object 4 is detected).
Since the parameter space is
4-dimensional, the simplex has 5 vertices. For each
move of the simplex, we plot the position of the vertex corresponding
to the largest value of the posterior distribution in the $(X,Y)$ and
$(A,R)$-subspaces; the position to which the simplex eventually
converged is indicated by the arrows.
After exploring the log-posterior
distribution for a few iterations, the simplex vertices are concentrated in
the neighbourhood of the global maximum. As the annealing temperature
is reduced the algorithm convergences ever more tightly on the
position of the peak.

With the choice of annealing schedule parameters $T_0$, $f$ and
$N_{\rm iter}$ discussed above, the detection of all the objects
listed in Table~\ref{tab6} required $\sim 40000$ evaluations of the
log-posterior. The entire analysis was completed
in $\sim 8$ mins on a Pentium III 1 GHz processor, which is somewhat
faster the {\sc McClean} algortihm, as might be expected.
Finally, we note also that 
the {\sc MaxClean} algorithm also obtains similar results to those 
produces by {\sc McClean} algorithm for the case in which the rms of the pixel
noise is decreased to 1 unit or increased to 3 units (see
section~\ref{mctoy}).

\section{Detecting the SZ effect in CMB maps}
\label{szdetect}

As a cosmological illustration of the Bayesian approach to 
detecting discrete objects in a background, in this section we consider the 
specific problem of identifying the Sunyaev-Zel'dovich (SZ) effect
from clusters embedded in maps of emission due to 
primordial cosmic microwave background (CMB) anisotropies. The
dominant microwave emission from clusters of galaxies is the thermal
SZ effect, in which CMB photons are inverse Compton scattered to
higher energies by fast moving electrons in the hot intracluster gas.
This leads to a characteristic frequency dependence for the thermal
SZ effect that differs markedly from that of the primordial CMB
emission. The kinetic SZ effect is due to the Doppler effect from
clusters with a peculiar velocity along the line of sight, and thus
has the same frequency behaviour as the primordial CMB emission.
Moreover, the kinetic SZ effect is typically an order of magnitude
smaller than the thermal effect.
A recent review of both the thermal and kinetic SZ effects is given by
Zhang, Pen \& Wang (2002).

The important problem of detecting SZ clusters in microwave maps dominated by
primordial CMB emission has been considered previously by several authors,
and most recently by Herranz et al. (2002a,b), who considered
the thermal SZ effect.
In these recent analyses, the identification and characterisation 
of thermal SZ clusters were performed using the scale-adaptive linear
filtering technique developed by Sanz et al. (2001). In the former,
only maps at a single observing frequency were considered, whereas in
the latter the different spectral dependences of the SZ and CMB
emission were also used to differentiate between the two (and other)
components.

The Bayesian approach to object detection 
outlined in the previous sections (using either
the {\sc McClean} or {\sc MaxClean} algorithms) can be
applied straightforwardly to either single or multi-frequency data.
In this section, however, we will concentrate on the analysis
of a single-frequency map. The Bayesian analysis of detailed simulated 
multi-frequency SZ and CMB observations will be presented in a
forthcoming paper. 

\subsection{The thermal SZ effect}

Let us first consider the detection of the thermal SZ effect in clusters.
Since our aim is merely to illustrate our Bayesian approach,
we shall analyse a simple set 
of simulated observations, at a single observing frequency, 
using the {\sc McClean} algorithm.
The simulations assume each SZ cluster to be spherically-symmetric 
with an electron density profile
of the form
\[
n_{\em e}(r) =
n_0\left[1+\left(\frac{r}{r_c}\right)^2\right]^{-3\beta/2}.
\]
In this case, one easily finds that the projected
thermal SZ profile on the sky, for a cluster 
centred at the position $\vect{X}=(X,Y)$ with a central SZ amplitude $A$,
is given by
\begin{equation}
\tau(\vect{x};X,Y,A,r_c) 
= A\left(1+\frac{|\vect{x}-\vect{X}|^2}{r_c^2}
\right)^{-\lambda}.
\label{king}
\end{equation}
where $\lambda=(3\beta-1)/2$. Assuming the standard value $\beta=2/3$
gives $\lambda=1/2$. In this case, $\tau(\vect{x})$ 
has the rather unpleasant feature of possessing an infinite volume, and 
so its Fourier transform is singular at the origin of 
$\vect{k}$-space. As advocated by Herranz
et al. (2002a), however, this troublesome behaviour
can be circumvented by replacing (\ref{king}), for the case $\lambda=1/2$, by
the modified profile
\begin{equation}
\tau(\vect{x};X,Y,A,r_c) 
= \frac{Ar_cr_v}{r_v-r_c}
\left(\frac{1}{\sqrt{r_c^2+d^2}}-\frac{1}{\sqrt{r_v^2+d^2}}\right),
\label{modking}
\end{equation}
where $d=|\vect{x}-\vect{X}|$ and $r_v$ 
may be interpreted as some typical `virial' radius for the cluster; 
we set $r_v=3r_c$. The profile (\ref{modking}) has a finite volume and 
a non-singular Fourier transform.

Our simulations are performed at an observing frequency of 100 GHz
on a $200\times 200$ grid, with a cell size
of 1.5 arcmin, so that the simulated sky area is
$5^\circ \times 5^\circ$.  At 100 GHz, the dominant sky emission is due to 
primordial anisotropies in the CMB and we may reasonably neglect contaminating
emission from the Galaxy. The CMB emission in our simulations is modelled
by a homogeneous Gaussian random field with
a standard inflationary cold dark matter power spectrum $C_\ell$.
The assumed cosmological model is spatially flat with $\Omega_{\rm b}=0.05$, 
$\Omega_{\rm cdm}=0.3$ and $\Omega_\Lambda=0.65$. We also assume a reduced
Hubble parameter $h=0.65$, a primordial scalar
spectral index $n=1$, and no tensor modes. The 
$C_\ell$ coefficients were created using {\sc Cmbfast}
(Seljak \& Zaldarriaga 1996). The rms of the CMB emission is $\sim$ 
100 $\mu$K.

To model the thermal SZ emission from clusters, we uniformly 
distribute 15 profiles of the form
(\ref{modking}) within the $5^\circ \times 5^\circ$ patch of sky. 
At 100 GHz, the thermal SZ effect produces
a decrement, and so we draw the central amplitude $A$ (in $\mu$K)
of each profile from the uniform distribution ${\cal U}(-500,-50)$. 
This corresponds SZ Compton $y$-parameters in the range $\sim 10^{-4}-
10^{-5}$, which is typical for
more realistic SZ simulations. The core radius
$r_c$ (in arcmin) of each cluster is drawn from the uniform distribution
${\cal U}(0.75,3)$.

\begin{figure*}
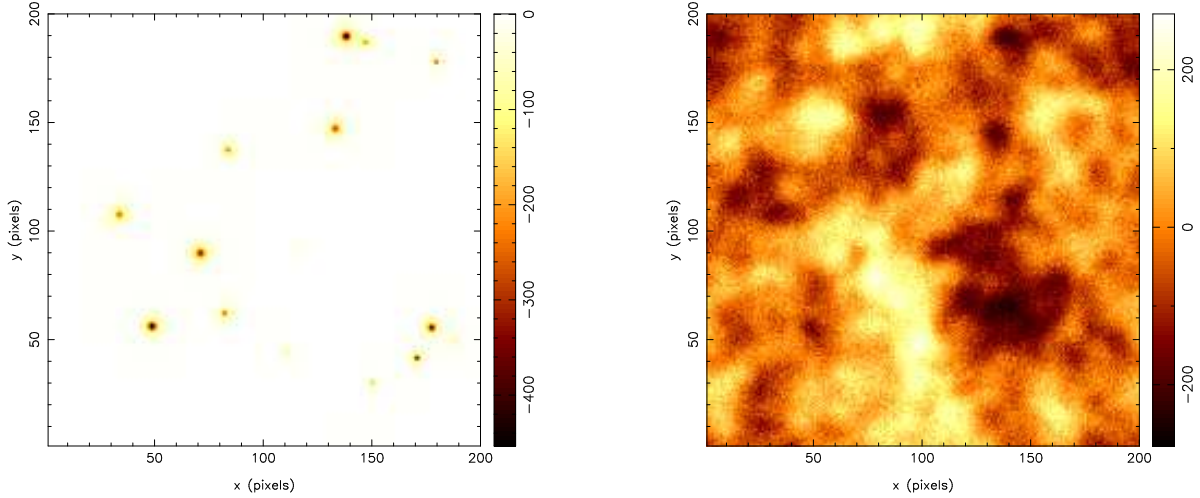

\begin{center}
\includegraphics[angle=-90,width=7cm]{fig7a.ps}
\qquad\qquad\qquad
\includegraphics[angle=-90,width=7cm]{fig7b.ps}
\caption{Left: the 15 thermal SZ modified King profiles used in our 
$5^\circ \times 5^\circ$ simulation (see
text for details). Right: simulated Planck data of the same
$5^\circ \times 5^\circ$ field at 100 GHz, assuming a 
Gaussian observing beam with a FWHM of 10 arcmin. In addition to the
convolved thermal SZ signal, the map contains
primordial CMB emission and instrumental noise (see text for details).
Both maps are plotted in units of $\mu$K.}
\label{fig7}
\end{center}
\end{figure*}

We include some experimental realism into our simulations by
performing simulated observations that correspond
to those expected from the forthcoming Planck mission. 
At 100 GHz, we model the Planck observing beam as a Gaussian 
with a FHWM of 10 arcmin, and we assume Gaussian instrumental pixel noise
which is homogeneous and uncorrelated and has an rms of 20 $\mu$K.
This noise level might reasonably be expected for the HFI 100 GHz channel 
after 24 months of obervation by Planck.
After convolution with the
beam, the central SZ amplitudes are considerably reduced and lie in
the range $-130$ to $-10$ $\mu$K. Comparing these amplitudes with the 
total rms of the convolved CMB emission and the instrumental noise, the
signal-to-noise ratio varies from 1.5 to 0.1, and 
this therefore presents a considerably challenging object detection
problem.
Fig.~\ref{fig7} shows the true thermal SZ emission in the simulations
(left panel) and the convolved noisy data, which is clearly dominated by
primordial CMB emission (right panel). 

\subsubsection{Evaluation of the posterior distribution}
\label{evalpost}

In order to perform the Bayesian object detection (using either
{\sc McClean} or {\sc MaxClean}), one needs to calculate
the posterior distribution a large number of times, each at a different
point in the parameter space $\vect{a}=\{X,Y,A,r_c\}$. Although 
the calculation of the priors is
trivial, the evaluation of the likelihood function is
computationally more demanding. Since the
background `noise' $\vect{n}$ 
(which consists of primordial CMB and instrumental
pixel noise) is a Gaussian random field, the likelihood function has
the form discussed in section~\ref{bayesobj}, namely
\begin{equation}
\Pr(\vect{D}|\vect{a}) = 
\frac{\exp
\left\{-{[\vect{D}-\vect{s}(\vect{a})]}^{\rm t}\vect{N}^{-1}
[\vect{D}-\vect{s}(\vect{a})]\right\}}
{(2\pi)^{N_{\rm pix}/2}|\vect{N}|^{1/2}}.
\label{reallike}
\end{equation}
Unlike the toy problem discussed in section~\ref{toyprob}, however, 
the presence of the correlated CMB emission means that
the `noise' covariance matrix $\vect{N}$ is not diagonal. Since its
dimensions are $N_{\rm pix}\times N_{\rm pix}$ (i.e. $200^2 \times
200^2$ for our simulations), the direct calculation of the quadratic
form and the determinant in (\ref{reallike}) would be an extreme computational
burden. Moreover, even the calculation of the signal
$\vect{s}(\vect{a})$ resulting from a source characterised by a given set of
parameters $\vect{a}$, would require one to perform a beam 
convolution for each likelihood evaluation. In general this requires
requires two Fast Fourier Transforms (FFTs), which are themselves
computationally intensive.

Fortunately, for the problem at hand, there exists a straightforward solution
to these computational difficulties, which is simply to perform the
entire calculation in Fourier space. Thus, we instead consider our
data to be the Fourier transform $\widetilde{\vect{D}}$ of our original data
map. Then, since the background `noise' is a 
homogeneous Gaussian random field, the likelihood function now takes the
form
\begin{equation}
\Pr(\widetilde{\vect{D}}|\vect{a}) = 
\frac{\exp
\left\{-{[\widetilde{\vect{D}}-\widetilde{\vect{s}}(\vect{a})]}^\dagger
\widetilde{\vect{N}}^{-1}
[\widetilde{\vect{D}}-\widetilde{\vect{s}}(\vect{a})]\right\}}
{(\pi)^{N_{\rm pix}}|\widetilde{\vect{N}}|},
\label{complike}
\end{equation}
where we note that several factors of two differ from the expression 
(\ref{reallike}),
since (\ref{complike}) is a multivariate Gaussian distribution of complex
variables (Eaton 1983). In this expression, since the `noise'
background is a homogeneous random field, the `noise' covariance matrix 
$\widetilde{\vect{N}}$ is diagonal. 
Indeed, if we denote the Fourier transform of
the observing beam by $\widetilde{B}(k)$, and the power spectra of the
primordial CMB and instrumentional noise by $P_{\rm cmb}(k)$ and
$P_{\rm n}(k)$ respectively, then
\begin{equation}
\widetilde{N}_{ij} 
 =  \langle
\widetilde{n}(\vect{k}_i)\widetilde{n}^\ast
(\vect{k}_j) \rangle 
 =  \left[|\widetilde{B}(k_i)|^2P_{\rm cmb}(k_i)+P_{\rm
n}(k_i)\right]\delta_{ij},
\label{ftncov}
\end{equation}
where $k=|\vect{k}|$. For the case of a Gaussian
observing beam with dispersion $\sigma_b$, the beam Fourier transform
has the form $\widetilde{B}(k) = \exp(-\sigma_b^2 k^2/2)$.

\begin{figure}
\begin{center}
\includegraphics[angle=-90,width=7cm]{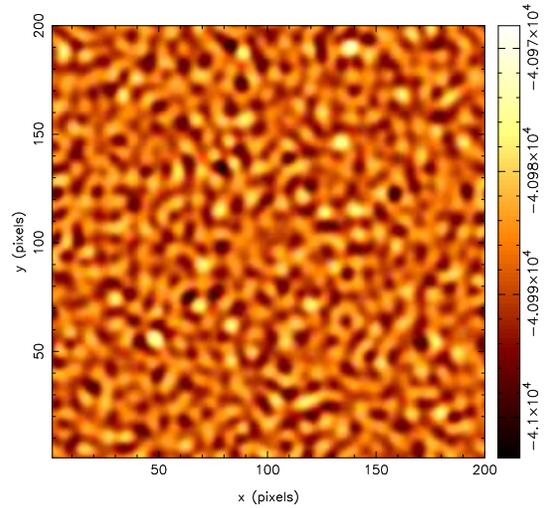}
\caption{The 2-dimensional conditional log-posterior distributions 
in the $(X,Y)$-subspace for the thermal SZ detection problem illustrated in
Fig.~\ref{fig7}. The values of the 
central SZ amplitude and core radius 
are conditioned at $A=-250$ $\mu$K and $r_c=1.5$ arcmin.}
\label{fig8}
\end{center}
\end{figure}
\begin{figure*}
\begin{center}
\includegraphics[width=14cm]{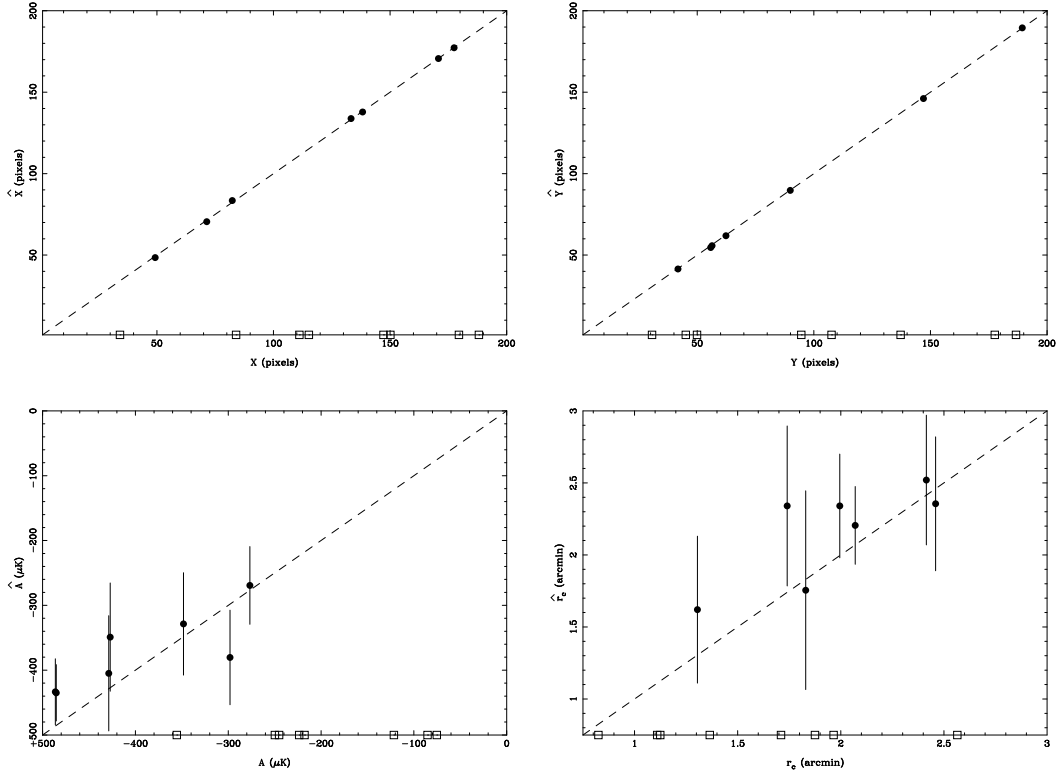}
\caption{Comparison of the true and estimated values of the
parameters $X$, $Y$, $A$ and $r_c$ for the 7 thermal SZ clusters
identified by the {\sc McClean} algorithm (filled circles). The
error bars on the estimated values denote the 68 per cent
confidence limits of the corresponding marginalised posterior
distribution. The 8 unidentified clusters are represented by the
open squares plotted along the bottom of each graph.}
\label{fig9}
\end{center}
\end{figure*}

The only remaining task in evaluating (\ref{complike}) is to calculate the
signal $\widetilde{\vect{s}}(\vect{a})$ resulting from a
discrete object of the form (\ref{modking}), characterised by a given set of
parameters $\vect{a} = \{X,Y,A,r_c\}$. Fortunately, the
two-dimensional Fourier transform of (\ref{modking}) is easily calculated
to be
\begin{equation}
\widetilde{\tau}(\vect{k};X,Y,A,r_c) = A\exp(-i\vect{k}\cdot\vect{X})
\widetilde{\tau}(k;0,0,1,r_c),
\label{ftobj}
\end{equation}
where $\vect{X}=(X,Y)$ and the
Fourier transform $\widetilde{\tau}(k;0,0,1,r_c)$
of the `standard object' is given by
\begin{equation}
\widetilde{\tau}(k;0,0,1,r_c)=\frac{2\pi r_vr_c}{r_v-r_c}
\frac{\exp(-r_c k)-\exp(-r_v k)}{k}.
\label{ftmodking}
\end{equation}
Despite appearances to the contrary $\widetilde{\tau}(k;0,0,1,r_c)$
is well-defined at the origin of Fourier space, as is easily seen by
evaluating the limit $k \to 0$. The signal $\widetilde{s}(\vect{a})$
in the data produced by an object with a particular 
parameterisation is then obtained by simply multiplying (\ref{ftobj}) 
by the beam Fourier transform $\widetilde{B}(k)$.

Thus, using (\ref{ftncov})--(\ref{ftmodking}), we may
calculate the likelihood function (\ref{complike}) at minimal computational
cost, entirely in Fourier space. Since we are assuming uniform priors
on each of the parameters $\{X,Y,A,r_c\}$, the evaluation of the
posterior distribution at any point in parameter space is also
straightforward. To illustrate the structure of this function, 
in Fig.~\ref{fig8} we plot the
2-dimensional conditional log-posterior distribution in the 
$(X,Y)$-subspace corresponding to the simulated data shown in
Fig.~\ref{fig7}; in this plot the
values of $A$ and $r_c$ are conditioned at $A=-250$ $\mu$K and $r_c=1.5$
arcmin (i.e. 1 pixel). We see that faint local maxima do indeed
occur at the positions of the true SZ clusters, although these
peaks are not greatly pronounced with respect to `chance' fluctuations in the
log-posterior caused by the CMB emission and the instrumental noise.
This problem therefore represents a severe challenge for any object
detection algorithm.

\subsubsection{Application of the {\sc McClean} algorithm}
\label{tszapp}

We now attempt to recover the SZ cluster profiles in the simulated
data using the {\sc McClean} algorithm discussed in section~\ref{mccleansec},
which identifies objects iteratively by
MCMC sampling from the full 4-dimensional posterior distribution.
As in the toy problem, we use 5 Markov chains in the sampling
process, and take 500 post burn-in samples from the posterior at
each iteration.

The algorithm identifies 7 of the 15 objects present, before terminating when 
the evidence for a further object falls below that for the model 
containing no objects. The results are summarised in Fig.~\ref{fig9},
in which the 7 identified clusters are denoted by the solid circles, and
the open squares represent the 8 unidentified clusters.
We see that the positions of the identified clusters are very
accurately constrained. Indeed the error-bars plotted on the points are
not visible. The typical error in $X$ or $Y$ is $\sim 0.75$ arcmin
(i.e. $\sim 0.5$ pixels). The amplitude $A$ and core radius $r_c$ of
each identified cluster are also recovered to reasonable accuracy,
with the scatter about the true values well-described by the
derived error-bars. As one might expect,
the 7 clusters identified by the algorithm are those with
the largest central SZ decrements, except for one cluster with
$A=-355$ $\mu$K and $r_c=1.1$. This cluster is not detected since, 
owing to its small core radius, the
central amplitude of this cluster is considerably reduced by the
beam convolution. We note in particular that there are no spurious
detections. The entire analysis required around $2.5\times 10^5$
function evaluations and took $\sim 50$ mins of CPU time on
a Pentium III 1 GHz processor.

The above results assume an instrumental pixel noise rms of
20 $\mu$K, which corresponds to around 24 months of observation
for the Planck HFI 100 GHz channel. It is expected, however, that
preliminary Planck channel maps will be available after just 6 months
of observation. For the HFI 100 GHz channel, the rms of the 
instrumental pixel noise on such a map will thus be about 
$\sqrt{4} \times 20 = 40$ $\mu$K. Since it is hoped that a preliminary
catalogue of SZ clusters might be produced from these data, 
we have therefore repeated our analysis for this higher noise level. 
In this case, we find that the {\sc McClean} algorithm identifies the
same first five clusters that it found in the lower noise simulation, but
then terminates. Despite detecting fewer clusters, we find that the 
typical errors on the position, amplitude and
core radius of each identified cluster are very similar to those
obtained for the lower noise case, and so are not plotted here.

\subsubsection{Comparison with linear filter techniques}

It is worthwhile to compare the performance of the {\sc McClean}
algorithm with more traditional filtering techniques, such as those
proposed by Herranz et al. (2002a). In order to make a direct
comparison, we now apply the {\sc McClean} algorithm
to an alternative simulated dataset, which has the same statistical 
properties as `simulation 2' in Herranz et al. 

\begin{figure}
\begin{center}
\includegraphics[angle=-90,width=7cm]{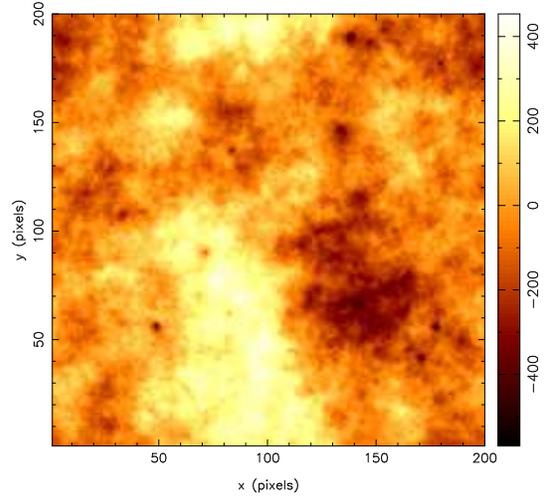}
\caption{Simulated map of a $5^\circ \times 5^\circ$ field at 100
GHz. The map consists
of the 15 modified King profile SZ clusters shown in Fig.~\ref{fig7}
(left panel) and a `CMB' signal with a $C_\ell \propto
\ell^{-3}$ power spectrum and an rms amplitude of $140$ $\mu$K.
The map units are $\mu$K.}
\label{fig9+1}
\end{center}
\end{figure}

Our simulation again consists of a $5^\circ \times 5^\circ$ patch of
sky, containing $200\times 200$ pixels of size 1.5 arcmin.  The
thermal SZ emission is modelled with the same 15 modified King profile
clusters as shown in Fig.~\ref{fig7} (left panel). As discussed
earlier, the central amplitudes of these SZ clusters are drawn from a
uniform distribution covering one order of magnitude in temperature
and their core radii are drawn uniformly between 0.75 and 3
arcmin. Indeed, this is precisely the statistical content of the SZ
signal assumed in `simulation 2' by Herranz et al.  The only
difference between the two simulations is the size of the patch of
sky. In `simulation 2', a $12^\circ.8\times 12^\circ.8$ patch of sky
was considered, consisting of $512\times 512$ pixels, in which 100
such SZ clusters were randomly distributed. Clearly, the two
simulations share the same pixel size and the same average number of
clusters per deg$^2$. Following `simulation 2', we add to the SZ
clusters a `CMB' signal consisting of a Gaussian random field with a
$C_\ell \propto \ell^{-3}$ power spectrum and an rms amplitude such
that the peaks of the SZ clusters are on average at the
$2\sigma$-level of the CMB signal.  The mean central amplitude of the
SZ clusters in our simulation is $-278$ $\mu$K, and so the rms of our
CMB signal is set to $140$ $\mu$K.  Again following `simulation 2', we
do {\em not} convolve the map with any observing beam, {\em nor} do we
add any instrumental pixel noise.  The resulting `data' map is shown
in Fig.~\ref{fig9+1}, for which the random seed variable used to
create the CMB emission is the same as that use to create the CMB
emission in Fig.~\ref{fig7} (right panel), thereby ensuring a similar
morphology for the two CMB fields.

\begin{figure*}
\begin{center}
\includegraphics[width=14cm]{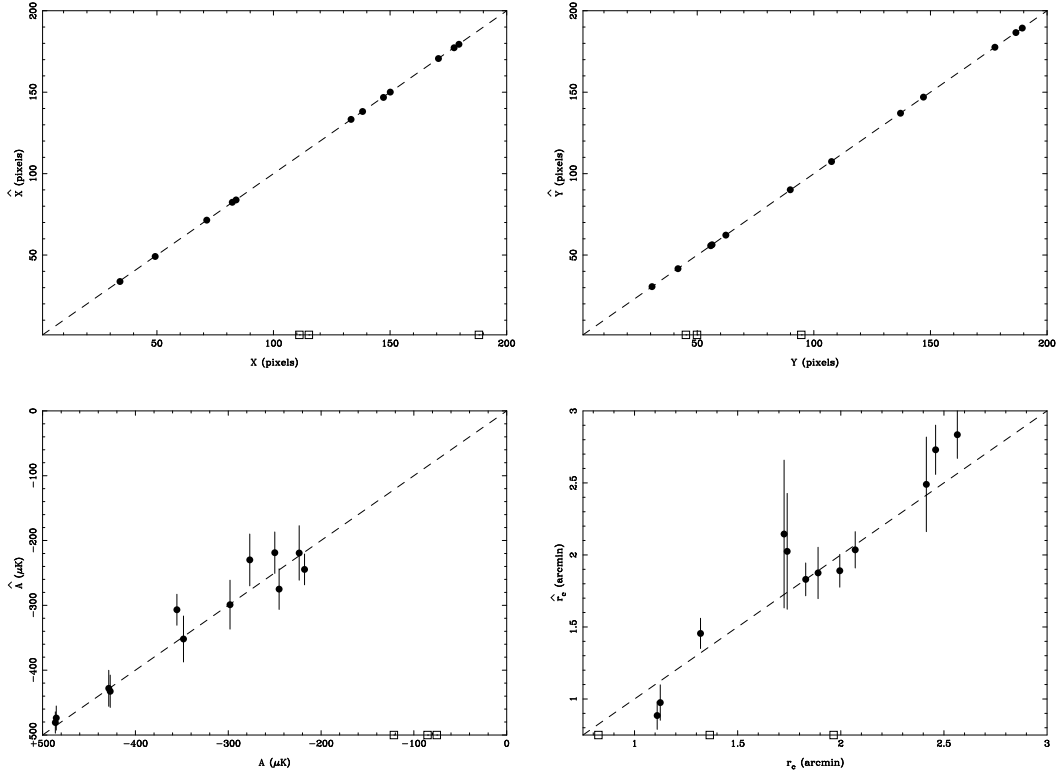}
\caption{Comparison of the true and estimated values of the
parameters $X$, $Y$, $A$ and $r_c$ for the 12 thermal SZ clusters
identified by the {\sc McClean} algorithm (filled circles). The
error bars on the estimated values denote the 68 per cent
confidence limits of the corresponding marginalised posterior
distribution. The 3 unidentified clusters are represented by the
open squares plotted along the bottom of each graph.}
\label{fig9+2}
\end{center}
\end{figure*}

We now apply the {\sc McClean} algorithm to this simulated map, again
using 5 Markov chains and taking 500 post burn-in samples from the
posterior at each iteration. In this case, the algorithm detects 12 of
the 15 objects before terminating. The results are summarised in
Fig.\ref{fig9+2}, in which the 12 identified clusters are denoted by
the solid circles, and the open squares represent the 3 unidentified
clusters.  We see that the positions of the identified clusters are
once again very accurately constrained. Also, the amplitude $A$ and core
radius $r_c$ of each identified cluster are recovered to
reasonable accuracy, with the scatter about the true values
well-described by the derived error-bars. We see that the 3
unidentified clusters are those with the smallest central amplitudes.
Moreover, we note that, once again, there are no spurious detections. 
The burn-in period required for
the Markov chains was somewhat shorter for this easier problem, and
the entire analysis took around $1$ hour of CPU time on
a Pentium III 1 GHz processor.

Let us now compare our results with those obtained by Herranz et
al., using their optimal linear pseudofilter and setting a
$4\sigma$ detection threshold above the background rms in the filtered domain.
Herranz et al. find that the 
linear filter identifies only 49 per cent of the SZ clusters present, 
whereas {\sc McClean} detects 80 per cent; neither method generates
spurious detections.  To compare the results more closely, we
calculate the `mean relative error' in the determination of
the cluster amplitudes and core radii. These are defined respectively
as 
\[
\bar{e}_A = \frac{1}{N_{\rm d}} \sum_{i=1}^{N_{\rm d}}
\frac{|\hat{A}_i-A_i|}{A_i}\quad\mbox{and}\quad
\bar{e}_{r_c} = \frac{1}{N_{\rm d}} \sum_{i=1}^{N_{\rm d}}
\frac{|\hat{r}_{c,i}-r_{c,i}|}{r_{c,i}},
\]
where $N_{\rm d}$ is the number of clusters detected. For the linear
filter, Herranz et al. found $\bar{e}_A=0.104$ and
$\bar{e}_{r_c}=0.140$, whereas the {\sc McClean} algorithm gives
$\bar{e}_A=0.059$ and $\bar{e}_{r_c}=0.096$. Thus, we conclude that,
on average, {\sc McClean} determines the amplitude and core radius of
the detected clusters with an accuracy almost twice that of the 
linear filter. Moreover, the Bayesian approach also yields reliable
error estimates on the derived parameters.

\subsection{The kinetic SZ effect}

As stated earlier, the kinetic SZ effect is typically an order of
magnitude smaller than the thermal effect. Thus its detection in
microwave maps dominated by primordial CMB emission presents an
extreme challenge. We note that an optimal linear filtering technique 
for detecting the kinetic SZ
effect embedded in maps of primordial CMB emission
has been discussed by Haehnelt \& Tegmark (1996).

We shall again consider
a simple set of simulated observations, very simliar to
those considered in section~\ref{tszapp}. 
Since we are restricting ourselves in this
paper to the analysis of simulated observations at a single frequency,
we shall consider simulated Planck observations at 217 GHz. At this 
frequency, the emission due to the thermal SZ effect is zero, and so
can be ignored. Moreover, this channel is still in the favourable
frequency range where the diffuse Galactic foreground emission
is small. Thus, one need only consider the kinetic SZ effect
from clusters, embedded in primordial CMB emission (and instrumental
noise). We model the kinetic SZ effect using the 
same 15 cluster profiles as used above for the thermal effect, but
now with the central amplitude (in $\mu$K) drawn from the uniform
distribution ${\cal U}(-50,50)$. Once again, we assume primordial
CMB emission with rms $\sim 100$ $\mu$K. We model the Planck observing
beam at 217 GHz by a Gaussian with a FWHM of 5 arcmin, and we assume
uncorrelated instrumental pixel noise with rms $20$ $\mu$K.

\subsubsection{Evaluation of the posterior distribution for each cluster}

Since the kinetic SZ effect is of such small amplitude as compared
with the primordial CMB emission, we cannot hope to detect the
clusters in the same straightforward way as we did for the thermal
effect. Nevertheless, a reasonable strategy does present itself.
Since we have already detected and characterised the thermal
SZ emission in 7 of the clusters, we can assume the
position $(X,Y)$ and core radius $r_c$ obtained for each of these clusters
and attempt to recover only the amplitude $A$ of the kinetic
effect in each cluster individually.

By adopting this strategy, we effectively reduced our problem
to determining the 1-dimensional posterior distribution
$\Pr(A|\widetilde{D},\hat{X},\hat{Y},\hat{r}_c)$. In this case there is no need
to perform MCMC sampling from the posterior or to locate its
global maximum using a simulated annealing approach. Since it is 
1-dimensional, one can efficiently evaluate the posterior directly at
a regular set of points along the $A$-axis. These posterior
distributions can then be used to obtain a `best' estimate $\hat{A}$ of the 
central kinetic SZ amplitude in each cluster (for example, the peak or
the mean of the posterior), and to place confidence limits on our
estimate.

\begin{figure*}
\begin{center}
\vspace{7mm}
\includegraphics[width=7cm]{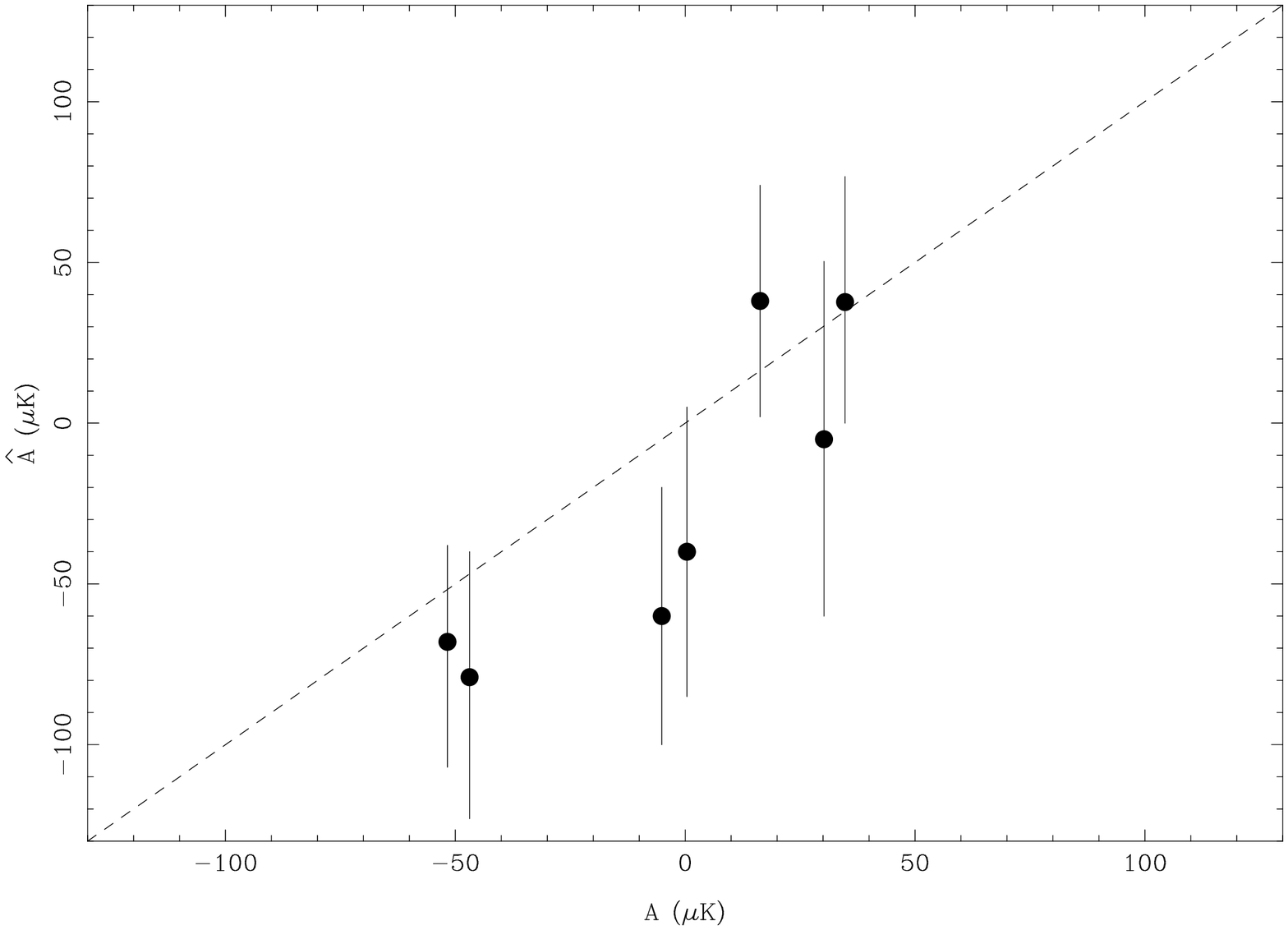}
\qquad\qquad\qquad
\includegraphics[width=7cm]{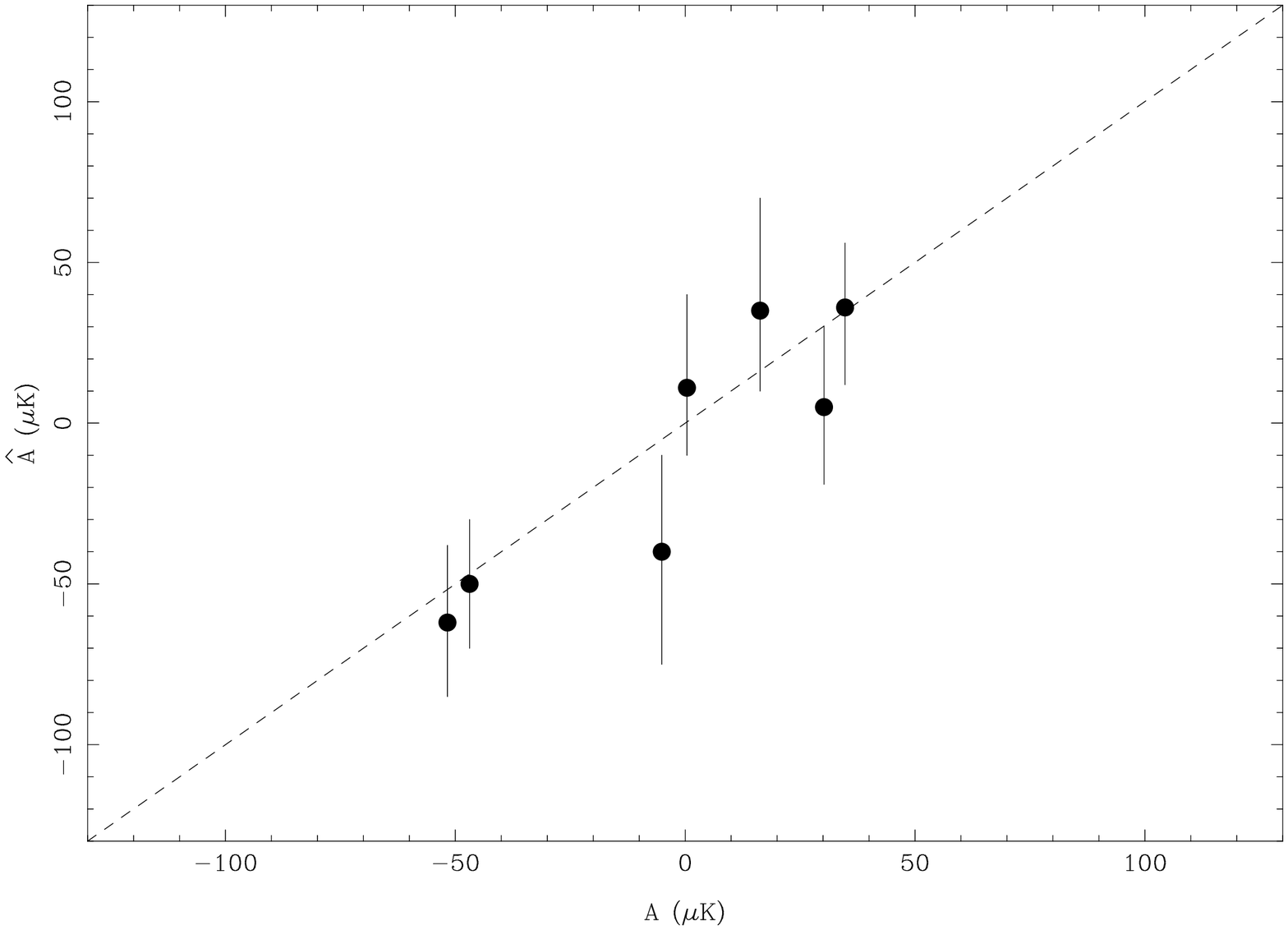}
\caption{Comparison of the true and estimated values of the
central kinetic SZ amplitude $A$ for the 7 clusters identifed earlier
by their thermal SZ effects, obtained from simulated 217GHz 
Planck data with a noise rms of 20 $\mu$K (left panel) and
5 $\mu$K (right panel). The solid circles denote the mean
of the 1-dimensional posterior distribution for $A$ for each cluster,
whereas the error-bars represent the 68 per cent confidence limit in
each case.}
\label{fig10}
\end{center}
\end{figure*}

We perform this procedure for the 7 clusters identified above from their
thermal SZ effect. For each cluster, the evaluation of the
1-dimensional posterior for $A$ requires only a few seconds of CPU
time. The results are plotted in Fig.~\ref{fig10} (left
panel), in which the solid circles represent the mean of the posterior
for each cluster and the error-bars represent the 68 per cent
confidence limits.
The left panel shows the results obtained for 
a realistic instrumental noise level of 20 $\mu$K per pixel, whereas the
right panel corresponds to an artificially low noise level of
5 $\mu$K per pixel. In the former, we see that the recovery
central kinetic SZ amplitudes is quite poor. The estimated
amplitudes do follow the true amplitudes, but 
with a large scatter. Indeed, the typical error in the recovered amplitude
is $\sim 50$ $\mu$K. In the lower noise case, however, the recovery
is better, with a typical error of $\sim 25$ $\mu$K. Even in this
extremely optimistic case, however, using the kinetic SZ effect to determine 
bulk flows in the Universe will be a challenging
task for the Planck mission.

\subsubsection{Comparison with linear filtering techniques}

We may straightforwardly compare our results with those obtained
using the linearly-optimal Fourier-filtering technique proposed by
Haehnelt \& Tegmark (1996) for recovering
the kinetic SZ effect in microwave maps 
dominated by primoridial CMB anisotropies. 

For a typical cluster, the
central kinetic SZ amplitude is
\begin{equation}
A \approx 30 
\left(\frac{n_{\rm e}}{3\times 10^{-3}~\mbox{cm$^{-3}$}}\right)
\left(\frac{R_{\rm c}}{0.4~\mbox{Mpc}}\right)
\left(\frac{v_{\rm p}}{500~\mbox{km s$^{-1}$}}\right)~\mbox{$\mu$K},
\label{typclust}
\end{equation}
where $n_{\rm e}$ is the electron density in the core, $R_{\rm c}$ is
the core radius and $v_{\rm p}$ is the radial peculiar velocity of the
cluster. This corresponds to a central optical depth due to electron
scattering for the cluster of $\tau \approx 0.005$.
Haehnelt \& Tegmark 
show (in their Fig.~3a)
that, for an instrumental noise rms of $\sim 6$ $\mu$K per 5 arcmin
beam (which is equivalent to 20 $\mu$K per 1.5 arcmin pixel) and
projected cluster core radii in the range 1--3 arcmin, the
$1\sigma$ error in determining the peculiar velocity of the cluster
is
\[
\Delta v_{\rm p} \approx 400 \left(\frac{\tau}{0.02}
\right)^{-1}~\mbox{km s$^{-1}$}.
\]
Thus, for a typical cluster with $\tau \approx 0.005$, one obtains
$\Delta v_{\rm p} \approx 1600$ km s$^{-1}$. From (\ref{typclust}),
this translates into an uncertainty in the central kinetic SZ
amplitude of $\Delta A \approx 100$ $\mu$K. Comparing this value with
the typical error $\Delta A \approx 50$ $\mu$K obtained by {\sc
McClean}, we see that the Bayesian approach is again around twice as
sensitive as the optimal linear filter.
  
\subsection{Future analysis of real {\sc Planck} data}

So far, we have considered the simple case of detecting
circularly-symmetric SZ clusters (with a simple modified King profile)
embedded in a background consisting of correlated (CMB) and
uncorrelated (instrumental noise) homogeneous Gaussian random fields,
observed at a single frequency. Clearly, this is a considerable
simplification of the situation one is likely to encounter in the
future analysis of real {\sc Planck} data. Several complications 
(or generalisations) need to be considered, of which the most
important are: (i) the Planck
instrument will observe at numerous different frequencies; (ii) at
some frequencies there will be significant non-Gaussian, 
non-stationary emission from Galactic foregrounds; (iii) the
instrumental noise will be non-stationary.

Let us first consider point (i). The presence of observations at
numerous frequencies brings both advantages and disadvantages to the
detection of SZ clusters. Clearly, the well-defined frequency spectrum
of the thermal SZ effect allows one to use multifrequency observations
to great effect in detecting clusters. Indeed,
techniques that make use of multifrequency data in this way
have already been proposed for detecting thermal SZ clusters in Planck data
(Diego et al. 2001, Herranz et al. 2002b). The extension of the
Bayesian approach presented here to multifrequency observations will
be presented in a forthcoming paper.

Multifrequency observations do, however, naturally lead to
consideration of point (ii) above, namely the presence of
foregrounds. The main theoretical difficulty in dealing with
foreground emission, such as Galactic synchrotron or dust emission, is
that, unlike the CMB, they are most likely to be non-Gaussian and
non-stationary. Similarly, any non-stationarity in the
instrumental noise field, as mentioned in point (iii) above, will
cause corresponding problems. 

From section~\ref{evalpost}, we see that the Bayesian method assumes
knowledge of the functional form of the likelihood function. Moreover,
in order to evaluate this function quickly in Fourier space, one 
assumes that the background (CMB, Galatic components and noise) is
stationary, so that its correlation structure can be described
entirely in terms of a power spectrum. It must also be noted, however,
that the same assumptions are required in the derivation of the optimal
linear filters proposed thus far. The assumption of stationary is
usually explicit (see e.g. Sanz et al. 2001), but by requiring
the variance of the filtered background to be minimised, the notion of
Gaussian distributed errors is also implicitly introduced.

For the Bayesian approach, the problem of non-Gaussianity can be
addressed directly by simply adopting the appropriate likelihood
function (if known). Nevertheless, even for highly non-Gaussian
fields such as a dust map, we find (as a result of the central limit
theorem) that the probability distribution of its Fourier modes is
reasonanly well-described by a Gaussian. We therefore expect the
non-Gaussianity of the background will not pose severe problems for either the
Bayesian or linear-filter object detection methods. Potentially more
serious is the non-stationary of the background. This leads to
correlations between Fourier modes, so that the covariance matrix
$\vect{N}$ in (\ref{complike}) is no longer diagonal. Although, in
principle, these correlations could be included explicitly (if known), the
evaluation of the resulting likelihood function would be
computationally more demanding (although not impossible). 
It has been demonstrated, however,
that neglecting inter-mode correlations does not significantly reduce
the accuracy of reconstructions in diffuse component separation
(see e.g. Stolyarov et al. 2002), and one would expect the same to be
true for discrete object detection. In any case, one may easily
investigate the effects non-stationarity and non-Gaussianity of the
background through simulations. As mentioned above, 
a detailed account of the issues associated with 
identifying SZ clusters in multifrequency simulated Planck observations
will be presented in a future paper.

\section{Conclusions}
\label{conc}

In this paper, we have presented a Bayesian approach for detecting and
characterising the signal from discrete objects embedded in a diffuse
background, which is a generic problem in many areas of astrophysics
and cosmology. Although our general approach is
applicable to datasets of arbitrary dimensionality, we have illustrated
our discussion by focussing on the important specific
example of detecting discrete objects in a diffuse background in a
two-dimensional image. 

The traditional approach to this problem has been to apply linear
filtering techniques. Typically the filter is optimised
for the detection of objects with a given spatial template, although
several different filters may be necessary if the objects vary
in size and shape. In these approaches, however, 
the distinction between the filtering and object-detection steps 
is somewhat arbitrary, and the
filtering process itself is only optimal among the rather limited class
of linear filters. 

The Bayesian approach presented in this paper 
also assumes a parametrised form for the objects of interest, but the
optimal values of these parameters, and their associated errors, are
obtained in a single step by evaluation of their 
full posterior distribution, given the data. 
If available, one may also place physical
priors on the parameters defining an object and on the number of
objects present. This technique provides the theoretically-optimal
method for parametrised object detection and characterisation.
Moreover, owing to its general nature, the
method may be applied to a wide range of astronomical datasets.

Within the Bayesian framework, two alternative
strategies are investigated: the simultaneous detection
of all the discrete objects in the image, and the iterative
detection of objects one-by-one. In both cases, the 
parameter space characterising the object(s) is explored
using Markov-Chain Monte-Carlo (MCMC) sampling. The specific
implementation of the MCMC technique used is that provided by
the {\sc Bayesys} software. We find that, although the simultaneous
detection of all objects present is the theoretically most desirable
approach, it
requires a complicated MCMC algorithm with the ability to sample
from subspaces of different dimensionality. 
We illustrate the technique by application 
to a toy problem in which Gaussian-shaped
objects are embedded in uncorrelated pixel noise. We find that
it performs well, but is also extremely computationally demanding.

In the iterative approach, one attempts to detect and characterise
the objects one-by-one. Thus at each iteration, the model for the
data contains just a single parametrised object. In this way, the
dimensionality of the parameter space is fixed and, generally, small. 
This allows much more rapid MCMC sampling from the posterior
distribution. Indeed, one is interested only in the shape of
the posterior distribution around its global maximum, and so fewer
samples are required. Once the dominant object has been detected
and characterised, it is subtracted from the data and the process
is repeated. The iterations are halted when the Bayesian evidence for
the `detected' object falls below the evidence for there being 
no objects left in the image. We call this approach the
{\sc McClean} algorithm. On applying this iterative algorithm to
the toy problem, we find that it performs equally well as the
simultaneous approach and is computationally much faster.

For the iterative detection
of objects, another approach is simply to locate the global maximum
of the posterior at each iteration using a simulated annealing simplex
technique. We call this algorithm {\sc MaxClean}. When applied to the
toy problem, we find that it yields results very similar to those
obtained by {\sc McClean}, and is computationally faster still.
This algorithm also has the advantage of being easily implemented
using only standard, readily-available numerical algorithms.

A cosmological illustration of the iterative approach is also
presented, in which the thermal and kinetic Sunyaev-Zel'dovich effects from
clusters of galaxies are detected in microwave maps dominated by emission
from primordial cosmic microwave background anisotropies.
We restricted our attention to the analysis of simulated 
single frequency data from the forthcoming Planck mission. In
particular, we found that using only simulated data from the 
HFI 100 GHz channel, the
{\sc McClean} algorithm is able to detect clusters with
central thermal SZ decrements $A \la -250$ $\mu$K and
core radii $r_c \ga 1$ arcmin.

Once the position and core radius of each cluster has been
determined from its thermal SZ signature, we investigate using
these values to place constraints on the kinetic SZ in the detected
clusters. Using simulated Planck observations at 217 GHz, where the
thermal effect is negligible, and assuming realistic instrumental
noise with an rms $\sim 20$ $\mu$K per pixel,
we found that the amplitude of
the kinetic effect can only be determined to an accuracy of
$\sim 50$ $\mu$K. Assuming a very optimistic noise 
rms of $\sim 5$ $\mu$K, this error can only be reduced to $\sim 25$ $\mu$K.

In general, it is unnecessary to restrict our attention to the
analysis of a single astronomical image, 
such as a map at a particular observing
frequency. In many cases, several different images may be available,
each of which contain information regarding the discrete object of
interest. For example, forthcoming CMB satellite missions, such as the Planck
experiment, should provide high-sensitivity, high-resolution
observations of the whole sky at a number of different observing
frequencies. Owing to the distinctive frequency dependence of the
thermal SZ effect, it is better to use the maps at all the observed
frequencies simultaneously when attempting to detect and characterise
thermal SZ clusters hidden in the emission from other astrophysical
components. The application of our Bayesian approach to more
realistic multi-frequency simulated Planck observations will be
presented in a forthcoming paper.

Finally, as mentioned in section~\ref{mccleansec}, the iterative
approach to object detection, exemplified by the {\sc McClean} 
and {\sc MaxClean} algorithms, is a compromise between using 
the fully Bayesian approach, in which the objects are identified
simultaneously, and the desire to 
obtain results in a reasonable amount of CPU time. 
As we have shown, this simple iterative approach may be successfully
applied to problems in which the objects of interest are 
well-localised and the data are such that the
determination of the characteristics of one object does not
significantly affect the derived properties of any other. In an 
astronomical context, this is clearly the case for well-separated SZ
clusters observed by experiments for which the beam is narrow and has
negligible side-lobes, such as the Planck mission. Since the iterative
approach does, however, represent only a convenient approximation to
the fully Bayesian simultaneous detection of objects, 
there will inevitably be (common) situations for which it will
perform less well. 

A simple astronomical example of such a case is
provided by interferometric observations, for which the synthesised
beam often has non-negligible sidelobes of large spatial extent. 
Using the SZ effect as an example, 
in this case the observed flux of a given cluster would contain 
contributions from other clusters lying in the
side-lobes of the synthesised beam. This would lead to the basic
iterative algorithm either over- or underestimating the flux of the 
cluster, depending on the positions and amplitudes of the other
clusters with respect to the sidelobe structure. Nevertheless, a
simple modification of the basic iterative algorithm easily addresses this
difficulty. Once a set of objects have been identified
and characterised by the {\sc McClean} (or {\sc MaxClean}) algorithm,
one can refine the estimates of the parameters of the objects
as follows. For each object in turn, one re-runs the MCMC 
(or global maximisation) algorithm
to explore the parameter space $\{X_k,Y_k,A_k,R_k\}$, but
this time including the other (fixed) derived objects in the model of the
data. This allows the defining parameters of the $k$th object
to be refined in the presence of the other detected objects $(k '\neq k)$. 
By repeating this procedure until convergence is obtained, one thus
obtains a {\em joint} optimal set of defining parameters for the
objects, thereby circumventing the problem identified above. The
application of this extended iterative procedure to simulated
interferometer observations of the SZ effect in clusters will be
presented in a forthcoming paper.

\subsection*{ACKNOWLEDGMENTS}
We thank John Skilling for providing the {\sc Bayesys}
software and Steve Gull for helpful advice regarding its use. The
authors also acknowledge NESTA for their financial support of this work.

\bsp 
\label{lastpage}
\end{document}